\documentclass[12pt,epsfig]{iopart}
\usepackage{epsf}
\usepackage{epsfig}
\usepackage{amssymb}

\begin{document}

\title[Fluctuations]{Fluctuations in glassy systems}
\vskip 10pt
\author{
Claudio Chamon$^{1}$ and Leticia F. Cugliandolo$^{2}$}
\address{
$^1$ Physics Department, Boston University,
\\
590 Commonwealth Avenue, Boston, MA 02215, USA
\\ 
$^2$Laboratoire de Physique Th{\'e}orique  et Hautes {\'E}nergies, Jussieu, \\
5\`eme {\'e}tage,  Tour 25, 4 Place Jussieu, 75252 Paris Cedex 05, France
\\
{\tt chamon@bu.edu, leticia@lpt.ens.fr}
}
\date\today

\begin{abstract}
We summarize a theoretical framework based on global
time-reparametrization invariance that explains the origin of dynamic
fluctuations in glassy systems. We introduce the main ideas without
getting into much technical details. We describe a number of
consequences arising from this scenario that can be tested numerically
and experimentally distinguishing those that can also be explained by
other mechanisms from the ones that we believe, are special to our
proposal. We support our claims by presenting some numerical checks
performed on the $3d$ Edwards-Anderson spin-glass. Finally, we discuss
up to which extent these ideas apply to super-cooled liquids that have
been studied in much more detail up to present.
\end{abstract}

\maketitle
\tableofcontents

\section{Why glasses? vs. universality in glassy dynamics}
\label{sec:intro}


It is common to encounter in nature systems that resist equilibration
with their environments and display what is called {\it
glassiness}. The name is derived from what we normally know as
glasses, an irregular array of silicon and oxygen atoms without
crystalline order, much as a liquid, but as hard as a
solid. The molecular diffusion within glasses is extremely sluggish,
slowing down by over 10 orders of magnitude as the temperature is
slightly decreased near the operationally defined glass transition
temperature. Hence, the term {\it glassiness} became generically
associated with very slow dynamics~\cite{glasses-gen}.

So, {\it why glasses?} The answer to this question has been the focus
of much research effort for long. It is certainly a rather
difficult question, and there have been a number of ideas lined up
for trying to understand how material systems become glassy. It is not
even clear whether in many systems there is a thermodynamic phase one
can call glass, or there is simply a dynamical crossover at low
enough temperatures. 

Do we need to fully answer {\it why glasses?} before we really
further our understanding of {\it glassy dynamics}? We take the point
of view that, by starting from the fact that glassy systems {\it
exist} (as nature presents us with concrete examples), we can then
attempt to characterize whatever possible universal properties there
are in glassy dynamics.

To make this statement clearer, let us turn to a question that
Anderson poses in the introduction of his {\it Concepts in solids}
text~\cite{Anderson}: {\it why solids?} This question, again, is a
rather complex one, and it is not untwined from the question {\it why
glasses?} if we focus on why a regular array of atoms, as opposed to a
random packing, forms in the first place. Even if one assumes that a
crystalline structure forms at low temperatures, a detailed
quantitative analysis of the energetics remains to be done so as to
determine if the packing is hexagonal, cubic, body or face centered,
{\it etc}. Nonetheless, if one {\it starts} from the {\it existence}
(as observed in nature) of a state with broken translational symmetry,
one can construct theories of lattice vibrations (and quantize it) and
of electronic band structure (and discern between insulators, metals
and semiconductors). Solid state physics starts from {\it the} solid.

The approach we review in this paper relies on a similar philosophy:
we do not claim any theory of the glass transition, and we do not
attempt to answer {\it why glasses?} with this particular
approach. Our theory does not allow us to make non-universal
predictions, such as what the glass transition temperature 
is (if one can be defined)
for a certain material, or whether the material
displays glassy behavior at all. We aim at understanding if there is a
set of principles, guided by symmetry considerations, that can be used
to understand certain {\it universal} aspects of glassy dynamics {\it
once the glass state is presented}. For example, glasses
age~\cite{aging}.  We thus expect that there exists a unified approach
to describe aging phenomena, and we seek some guiding principles,
based on dynamical symmetries, that could allow us to understand
universal properties in the aging regime, including the scaling of
spatial heterogeneities.

We propose that the symmetry that captures the universal aging
dynamics of glassy systems is the invariance of an effective dynamical
action under uniform reparametrizations of the time
scales~\cite{Chamonetal1}-\cite{Jaubertetal}.  Such type of invariance had
been known to exist since the early days in which the {\it mean-field
equilibrium dynamics} of spin-glasses was tried to be
understood~\cite{Sompo,equil-dynamics} and it was later encountered in
the better formulated {\it out of equilibrium} dynamic formalism
applied to the same systems~\cite{Cuku1,Cuku2}.  The invariance means that
in the asymptotic regime of very long times a family of solutions to
the equations of motion is found. This `annoyance', we claim, has
actually a physical meaning and implications in the fluctuating
dynamics of real glasses. In order to make this statement concrete, it
is necessary to show that the invariance also exists in finite
dimensional glassy models. With this aim we showed that global time
reparametrization invariance emerges in the long times {\it action} of
{\it short-range spin-glasses} assuming causality and a separation of
time scales~\cite{Chamonetal1}-\cite{Castilloetal2}.  Basically, the second
assumption amounts to start from a glass state, where one may claim
there is a separation between, roughly, a time regime where relaxation
is fast and another where relaxation is slow. The invariance can then
be used to describe dynamic fluctuations in spin-glasses and, we
conjectured, in other glassy systems as well.

Physically, the emergence of time reparametrization invariance can be
though of in the following way. The out of equilibrium 
relaxation of  glassy systems is well characterized by two-time 
functions, either correlations or linear responses. In the 
slow and aging regime they depend on two times and time-translation
invariance is lost. The proper measure of `time' inside the system
is the value of the {\it correlation itself}, and not the `wall
clock' in the laboratory. For instance, in a spin-glass the proper
measure of sample {\it age} is the spin-spin overlap or in 
particle systems it is the incoherent correlation function. {\it Age}
measures can fluctuate from point to point in the sample, what we
called heterogeneous aging, with younger and older pieces (lower and
higher values of the correlation) coexisting at the same values of the
two laboratory times.  The fact that the effective
dynamical action becomes invariant under global time
reparametrizations, $t\to h(t)$, everywhere in the sample, means that
the action weights the fluctuations of the proper ages, $C({\vec
r};t_1,t_2)$, directly, and the times $t_1$ and $t_2$ in the action are
just integrated over as dummy variables. To draw an analogy, in
theories of quantum gravity the space-time variables
$X_\mu(\tau,\sigma)$ are the proper variables, and the action is
invariant under conformal transformations of the world-sheet parameters
$\tau$ and $\sigma$.

So what does global time-reparametrization invariance symmetry
concretely teach us about observables in glasses? So far we discussed
a global symmetry or invariance with respect to uniform
time-reparametrizations. By looking at spatially heterogeneous
reparametrizations, we can predict the behavior of local correlations
and linear susceptibilities and the relations between them.  For
example, we predict that, after a convenient normalization that we
explain in the main text, the triangular relation between the local
coarse-grained correlations, $C(\vec r; t_1,t_2)$, $C(\vec r;t_2,t_3)$
and $C(\vec r;t_1,t_3)$, as a function of the intermediate time $t_2$,
$t_3<t_2<t_1$, at all spatial points, $\vec r$, should be identical to
the global triangular relation, in the asymptotic limit of very long
absolute times and delays between them and very large coarse-graining
linear length~\cite{Jaubertetal}. Different sites can be retarded or
advanced with respect to the global behaviour but they should all have
the same overall type of decay. Similarly, the relation between local
susceptibilities and their associated correlations should be identical
all over the sample~\cite{Castilloetal1,Castilloetal2} 
leading to a uniform effective temperature~\cite{Cukupe}.


 The purpose of this article is to describe our current
understanding
 of dynamic fluctuations (heterogeneities) in the
non-equilibrium
 relaxation of glassy systems arising from the
time-reparametrization
 invariance
scenario~\cite{Chamonetal1}-\cite{Jaubertetal},\cite{Bustingorryetal1}.
We illustrate it by
 presenting critical tests.

The
structure of this review is the following. In
Sect.~\ref{sec:time-rep-inv} we explain the origin of
the time-reparametrization invariance scenario.  We do not present
detailed derivations that were already published but
we
 aim at highlighting the main ideas behind the scene. In
Sect.~\ref{sec:consequences} we list several measurable consequences
of the theory. We discuss how they have been examined numerically
in different glassy models. The discuss here which consequences
could
 also be explained by other approaches and which, we believe,
are
 unique to our scenario. Finally, in Sect.~\ref{sec:discussion}
we
 discuss the scenario. Since this research project is not closed
yet,
 we present some proposals for numeric and experimental tests as
well
 as some ideas about analytic calculations that could help us
understanding the limitations of our proposal.

\section{Time reparametrization invariance}
\label{sec:time-rep-inv}

In this Section we explain how global time-reparametrization
invariance develops asymptotically in the aging regime of glassy
models.  For the sake of simplicity we focus on the classical
formalism and at the end of this section we mention the
modifications introduced by quantum fluctuations.

\subsection{Mean-field models -- dynamic equations}

Schematic models of spin-glasses, structural glasses and
ferromagnetic clean coarsening are encoded in the family of 
$p$-spin models defined by~\cite{KTW} 
\begin{equation} 
H=-\sum_{i_1 i_2 \dots i_p} J_{i_1 i_2 \dots i_p} 
s_{i_1} s_{i_2} \dots s_{i_p}
\label{eq:pspin}
\end{equation}
with quenched disordered exchanges distributed with the Gaussian law
$P(J_{i_1 i_2 \dots i_p})\propto e^{-p! \, J_{i_1 i_2 \dots
i_p}^2/(2N^{p-1})}$.  The dynamic variables $s_i$, $i=1,\dots,N$,
are of Ising type, $s_i=\pm 1$, or satisfy a global spherical
constraint $\sum_{i=1}^N s_i^2 = N$. $p$ is an integer parameter: with
$p=2$ and Ising variables one mimics spin-glasses, with $p>2$ and
Ising or spherical spins the phenomenology of structural glasses is
recovered, and with $p=2$ and spherical spins one describes
ferromagnetic domain-growth in clean systems.  The sum runs over all
$p$-uplets; for this reason these models are `mean-field' in the sense
that the saddle-point evaluation of the partition function or the
dynamic generating functional is exact in the thermodynamic limit,
$N\to\infty$. The Hamiltonian (\ref{eq:pspin}) also represents the
potential energy of a particle with position $\vec s=(s_1,\dots,s_N)$
on an infinite dimensional hypercube ($s_i=\pm 1$) or hypersphere
($\sum_{i=1}^N s_i^2 =N$).

Dynamics is introduced with a Langevin equation that represents the
coupling of the spins to an equilibrated thermal environment. Ising
spins are then replaced by soft variables by introducing a double-well
potential energy, $\sum_{i=1}^N V(s_i)$, with $V(s_i) = a
(s_i^2-1)^2$. The initial condition is usually chosen to be random
thus mimicking a rapid quench from infinite temperature to the working
temperature $T$.  In the limit $N\to\infty$ {\it exact}
Schwinger-Dyson equations couple the global correlation and instantaneous 
linear response
\begin{eqnarray}
&&NC(t,t_w) = \sum_{i=1}^N s_i(t) s_i(t_w) 
\; , 
\qquad 
NR(t,t_w) = \sum_{i=1}^N 
\left. 
\frac{\delta s_i(t)}{\delta h_i(t_w)}
\right|_{h=0}
\;  
\end{eqnarray}
($\sum_{i=1}^N s_i(t) =0$ for all $t$). The field $h_i$ couples linearly 
to the spin variables and, in general, we are interested
in perturbing fields that are uncorrelated
with the equilibrium states of the systems. 
It is not necessary to average over thermal noise or quenched 
disorder since these quantities do not fluctuate
in the out of equilibrium regime reached when the infinite volume
limit ($N\to \infty$) has been taken  at the outset. 
Here are in what follows times are measured from an origin that corresponds
to the quench to the working temperature. 
Our 
study applies then in the order of limits 
\begin{equation}
\lim_{t_w\to\infty} \lim_{N\to\infty}
\end{equation} 
in which the exact {\it causal} Schwinger-Dyson equations
for spherical models at times $t\geq t_w$ read~\cite{Cuku1,LesHouches}
\begin{eqnarray}
&&
(\partial_t-z_t) C(t,t_w) = 
\int_0^t dt' \; 
\Sigma(t,t') C(t',t_w) + 
\int_0^{t_w} dt' \; 
D(t,t') R(t_w,t') 
\; , 
\label{eq:C}
\\ 
&& 
(\partial_t-z_t) R(t,t_w) = 
\int_{t_w}^t dt' \; 
\Sigma(t,t') R(t',t_w) +\delta(t-t_w)
\; , 
\label{eq:R}
\end{eqnarray}
where the vertex, $D$, and self-energy, $\Sigma$,  
are functions of 
$C$  and $R$
\begin{eqnarray} 
D(t,t_w) = \frac{p}{2} C^{p-1}(t,t_w)
\; , 
\;\;\;\;\;\;
\Sigma(t,t_w) = \frac{p(p-1)}{2} C^{p-2}(t,t_w) \, R(t,t_w)
\; . 
\label{eq:self-energy}
\end{eqnarray}
The Lagrange multiplier $z_t$ is fixed by requiring $C(t,t)=1$.  
For Ising problems the soft spin constraint can be treated in 
the mode-coupling approximation~\cite{Chke} or else, 
one can apply a $T-T_c$ expansion taking advantage of the fact that 
the phase transition is of second order for $p=2$ (Sherrington-Kirkpatrick
or SK model)~\cite{Cuku2}.

In such fully-connected models all higher order correlations and
linear responses factorize and can be written as functions of these two-time
functions.  Self-consistent {\it approximate} treatments of
interacting particle models with realistic potentials, like the
mode-coupling approach, yield similar equations with the addition of a
wave-vector dependence that in the present context can also be
taken into account by considering models of $d$-dimensional random manifolds
embedded in $N\to\infty$ dimensional spaces~\cite{PhysicaA,LesHouches}.

\subsection{Structural glasses: the $p\geq 3$ cases}

We now focus on the $p\geq 3$ cases that mimic the structural
glass problem. We mention at the end of this subsection 
the Ising $p=2$ (SK)
spin-glass case that is not conceptually different but just
technically more involved. In Sect.~\ref{subsec:p=2} we
discuss the $p=2$ spherical problem that yields a mean-field
description of coarsening phenomena and is rather different from the
point of view of time-reparametrization transformations.

Equations (\ref{eq:C}) and (\ref{eq:R} are causal and be solved numerically
by constructing $C(t,t_w)$ and $R(t,t_w)$ from the initial instant $t=t_w=0$.
An analytic solution is possible in the asymptotic limit, as we discuss 
below. We first present the main features of $C$ and $R$ and we 
later explain how these are obtained from the asymptotic analytic solution.

\begin{figure}
\centerline{
\includegraphics[width=8cm]{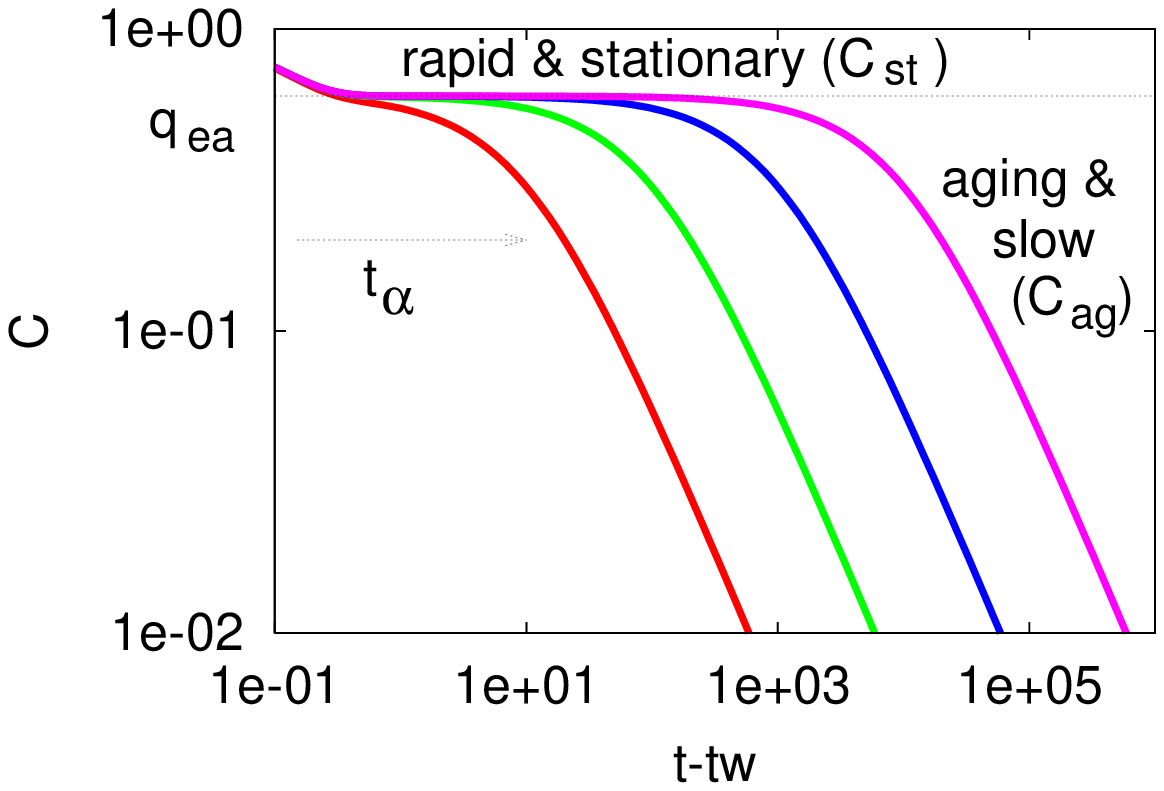}
\includegraphics[width=8cm]{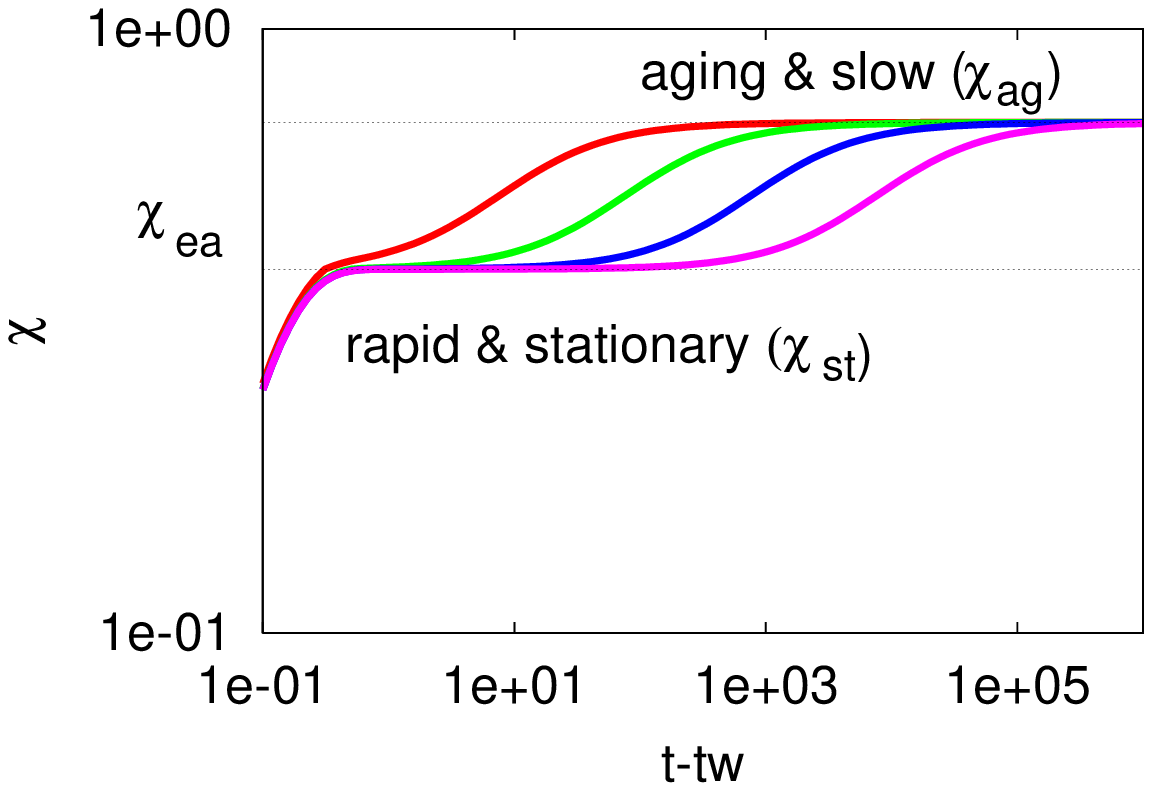}
}
\caption{Sketch of the relaxation of the self-correlation and
susceptibility in the glassy regime. The separation of time-scales is
clear in the figure. The Edwards-Anderson parameter, $q_{ea}$ the
corresponding susceptibility $\chi_{ea}$ and the asymptotic value
$\lim_{t\to\infty} \chi(t,t_w)$ are indicated with horizontal lines.}
\label{fig:sketch-fig}
\end{figure}

Equations~(\ref{eq:C}) and (\ref{eq:R}) have a dynamic transition
at a  critical temperature $T_d(p)$. At $T>T_d$ the  dynamics 
occurs in equilibrium and close to $T_d$ the decay of the 
correlations slows down as in super-cooled liquids with the $\alpha$ 
relaxation time diverging as a power law of $T-T_d$~\cite{KTW,Gotze}. 
Below $T_d$ 
eqs.~(\ref{eq:C}) and (\ref{eq:R}) admit a unique
solution~\cite{Cuku1,Cuku2} that is no longer stationary. The 
behaviour of the low-temperature correlation and susceptibility 
is sketched in Fig.~\ref{fig:sketch-fig}. 

The low-temperature solution
presents a separation of time-scales. In the long waiting-time limit
the self-correlation and integrated linear response or susceptibility,
$\chi(t,t_w) \equiv \int_{t_w}^t dt' \, R(t,t')$, can be written as
\begin{eqnarray}
C(t,t_w) &=& C_{st}(t-t_w) + C_{ag}(t,t_w)
\; , 
\\
\chi(t,t_w) &=& \chi_{st}(t-t_w) + \chi_{ag}(t,t_w)
\; . 
\label{eq:separation}
\end{eqnarray}
The first terms in the right-hand-side describe the stationary regime
at short time-differences in which the correlation and susceptibility
relatively rapidly approach a plateau at $\lim_{t-t_w\to\infty}
\lim_{t_w\to\infty} C(t,t_w) = q_{ea}$ and $\lim_{t-t_w\to\infty}
\lim_{t_w\to\infty} \chi(t,t_w) = (1-q_{ea})/T \equiv \chi_{ea}$.  The second
terms are the aging relaxation of $C$ towards zero (in the absence
of an external field), and the aging response of the system towards
the asymptotic value $\chi_{ea}+q_{ea}/T_{eff}$ with $T_{eff}$ a
parameter with the interpretation of an effective
temperature~\cite{Cukupe}.

The stationary and aging relaxation are {\it
fast} and {\it slow} in the sense that
\begin{eqnarray}
&
\partial_t C_{st}(t,t_w) \sim C_{st}(t,t_w) & 
\qquad\qquad\qquad C > q_{ea}
\; , 
\\
&
\partial_t C_{ag}(t,t_w) \ll C_{ag}(t,t_w) &
\qquad\qquad\qquad C < q_{ea}
\; ,
\label{eq:slowness-C}
\end{eqnarray}
and 
\begin{eqnarray}
&
\partial_t \chi_{st}(t,t_w) \sim \chi_{st}(t,t_w) & 
\qquad\qquad\qquad \chi < \chi_{ea}
\; , 
\\
&
\partial_t \chi_{ag}(t,t_w) \ll \chi_{ag}(t,t_w) &
\qquad\qquad\qquad \chi > \chi_{ea}
\; .
\label{eq:slowness-chi}
\end{eqnarray}
The aging self-correlation and susceptibility scale as
\begin{equation}
C_{ag}(t,t_w) 
\approx 
q_{ea} \; 
f_{\sc c}\left( \frac{{\cal R}(t)}{{\cal R}(t_w)}\right)
\; , 
\qquad 
\chi_{ag}(t,t_w) 
\approx 
q_{ea} \; 
f_{\chi}\left( \frac{{\cal R}(t)}{{\cal R}(t_w)}\right)
\; .
\label{eq:scaling-c}
\end{equation}
The scaling functions satisfy the limit conditions 
$f_c(1)=1$, $f_c(\infty)=0$, $f_\chi(1)=0$ and $f_\chi(\infty)=1/T_{eff}$.
Using mathematical properties of
monotonic two-time functions one can show that such a scaling holds
asymptotically in each (two) time-scale of the evolution~\cite{Cuku2}.
While in a system undergoing finite dimensional coarsening 
${\cal R}(t)$ has a natural interpretation as the {\it typical}
domain radius, in mean-field models there is no immediate
understanding of the `clock' ${\cal R}(t)$ that, in a sense, sets the
macroscopic time-scale.  The numerical solution suggests that ${\cal R}(t)$ 
is just a power of time.

\begin{figure}
\centerline{
\hspace{3.5cm}
\includegraphics[width=8cm]{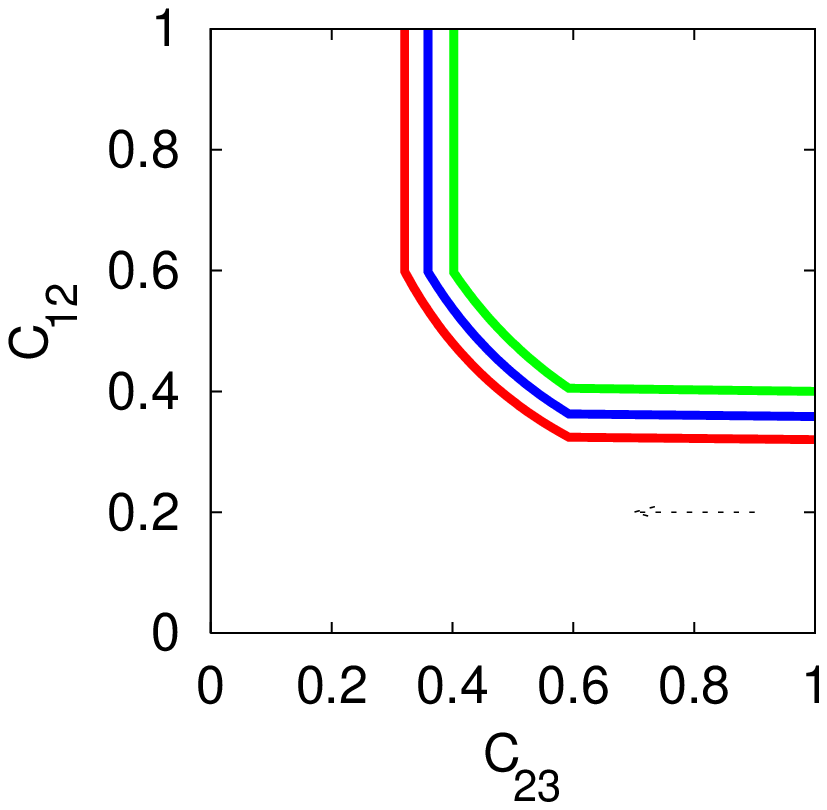}
\hspace{-1.5cm}
\includegraphics[width=8cm]{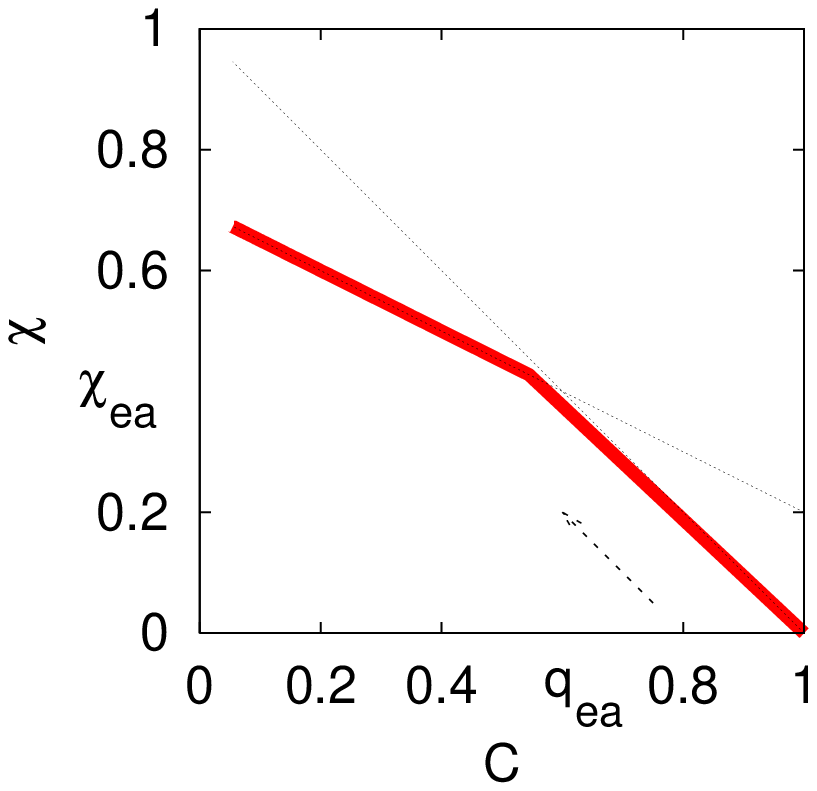}
}
\caption{Sketch of the parametric representation of the correlation
and susceptibility. Left: triangular relation between the
correlation function in the asymptotic limit $t_3\to\infty$. The three
curves correspond to different $t_1$'s such that $C(t_1,t_3)$ takes
three values. The breaking points lie at $q_{ea}$.  The arrow
indicates the sense of the evolution when $t_2$ increases from $t_3$
to $t_1$.
Right: susceptibility, $\chi(t_2,t_3)$ against correlation,
$C(t_2,t_3)$ at fixed $t_3$ using $t_2\geq t_3$ as a parameter in the
long $t_3$ limit.  The breaking point at $(q_{ea}, \chi_{ea})$
separates the stationary regime where the equilibrium {\sc fdt} is
satisfied from the aging regime where it is modified. The slopes are
$-1/T$ and $-1/T_{eff}$, respectively. The arrow also indicates here
the sense of the evolution when $t_2$ increases from $t_3$.}
\label{fig:sketch-fig2}
\end{figure}

In the asymptotic limit in which the additive separation of
time-scales with the scaling form (\ref{eq:scaling-c}) holds it is
convenient to use a parametric description of the dynamics in which
times do not appear explicitly. More precisely, the approach to the
asymptotic scaling, and $f_c$ and $f_\chi$, can be put to the test by
constructing `triangular relations' between correlations and
susceptibilities, respectively.  For generic three long times $t_1\geq
t_2\geq t_3\gg t_0$ one computes the correlations $C(t_\mu,t_\nu)$,
$\mu>\nu=1,2,3$. If the times are such that the ratios ${\cal
R}(t_\mu)/{\cal R}(t_\nu)$ remain finite in the asymptotic limit, that
is to say $C(t_\mu,t_\nu)=C_{ag}(t_\mu,t_\nu)$, one has
\begin{equation}
C(t_1,t_3) = q_{ea} \; 
f_c\left\{ 
f^{-1}_c[C(t_1,t_2)/q_{ea}] \; 
f^{-1}_c[C(t_2,t_3)/q_{ea}] 
\right\} 
\; . 
\label{eq:triangular} 
\end{equation}
If, instead, $t_1 = t_2 + \tau$ with $\tau>0$ finite, and ${\cal
R}(t_2)/{\cal R}(t_3)$ finite in such a way that $C(t_1,t_2)=
q_{ea} +C_{st}(t_1-t_2)$ and $C(t_2,t_3)=C_{ag}(t_2,t_3)$ one has
\begin{equation}
C(t_1,t_3) = \min \left[ \,C(t_1,t_2), C(t_2,t_3) \, \right] 
\label{eq:ultrametric} 
\end{equation}
asymptotically. 
This form goes under the name of {\it dynamic ultrametricity}.  In the
opposite case $t_3=t_2-\tau$ and ${\cal R}(t_1)/{\cal R}(t_2)$ finite
dynamic ultrametricity also holds.  These relations follow immediately
from the additive separation of time-scales (\ref{eq:separation}) and
the scaling (\ref{eq:scaling-c}) but they can be shown without
assuming dynamic scaling, just by using the monotonicity properties of
temporal correlations~\cite{Cuku2}.  The simplest way to see these
relations at work is to display $C(t_1,t_2)$ against $C(t_2,t_3)$, for
a chosen value of $C(t_1,t_3)<q_{ea}$, in a parametric plot in which
$t_2$ varies from $t_3$ to $t_1$. In the asymptotic limit
$t_3\to\infty$ the construction reaches a stable master curve as
displayed in Fig.~\ref{fig:sketch-fig2}-left.  The vertical and
horizontal parts correspond to $t_2$ such that $C(t_1,t_2)>q_{ea}$ and
$C(t_2,t_3)>q_{ea}$, respectively, and dynamic ultrametricity holds.  
The curved part is for $t_2$ such
that all correlations are in the aging regime and its functional form
is fully determined by $f_c$. The `clock' ${\cal R}$ yields the speed
at which the data-point moves on the parametric curve. A similar 
construction can be done for the susceptibilities. 

The stationary correlation, $C_{st}$, and susceptibility,
$\chi_{st}$, are linked by the equilibrium {\it fluctuation
dissipation theorem} ({\sc fdt}), $\chi_{st} = (1-C_{st})/T$. In the
aging regime, instead, there is a non-trivial relation between $\chi$
and $C$: $\chi_{ag}=(q_{ea}-C_{ag})/T_{eff}$ that yields
$f_\chi(x)=(1-f_c(x))/T_{eff}$. This relation is also better 
appreciated if shown in a parametric construction in which times
do not appear explicitly. In the long waiting-time limit 
the plot $\chi(t,t_w)$ against $C(t,t_w)$ with $t$ the parameter 
running from $t_w$ to infinity approaches a broken line
form with the slopes $-1/T$ (for $C>q_{ea}$) and $-1/T_{eff}$
(for $C<q_{ea}$). Again, 
the `clock' ${\cal R}$ yields the speed at which this 
curve is constructed upon increasing $t$.

An analytic solution to the Schwinger-Dyson equations was derived in
the limit of long waiting-time in which the separation of time-scales,
that is to say the plateaus in $C$ and $\chi$, are fully
established. In the aging regime, one uses the fact that the variation
of the correlation and linear susceptibility are negligible with
respect to all terms in the right-hand-side of the equations and can
thus be dropped.  Furthermore, one approximates the integrals by
separating the contributions from the stationary and aging
regimes~\cite{LesHouches}. Equation~(\ref{eq:R}) becomes
\begin{eqnarray}
\mu_\infty R_{ag}(t,t_w) \sim \frac{p(p-1)}{2}  
\int_{t_w}^t dt' \; C^{p-2}_{ag}(t,t') R_{ag}(t,t') \, R_{ag}(t',t_w)
\; 
\label{eq:slow-Req}
\end{eqnarray}
($\mu_\infty$ is a constant with contributions from $\lim_{t\to\infty}
z_t$ and border terms in the integrals). The companion
eq.~(\ref{eq:C}) takes a similar form within the same approximation.  Now,
the surprise is that the {\it approximate} equations are invariant
under the transformation
\begin{eqnarray}
t \to h_t \equiv h(t)
\; , \qquad\qquad
&&
\left\{ 
\begin{array}{l}
C_{ag}(t,t_w) \to C_{ag}(h_t,h_{t_w}) \; , \\ 
R_{ag}(t,t_w) \to \dot h_{t_w} \, 
R_{ag}(h_t,h_{t_w}) 
\; , 
\end{array}
\right.
\label{eq:rpg}
\end{eqnarray}
with $h_t$ positive and monotonic and $\dot h_{t_w}\equiv
dh_{t_w}/dt_w$.  While the functions $f_c$ and $f_\chi$ and their {\sc
fd} relation are fixed by the remaining approximate equation
(\ref{eq:slow-Req}) and its companion, the {\it time-reparametrization
invariance} does not allow one to compute, analytically, the clock
${\cal R}(t)$. This problem is similar to the velocity {\it selection}
problem present for instance in Fisher differential equation
describing front propagation and the like~\cite{Fisher}. The exact
Schwinger-Dyson equations {\it do} have a {\it unique} solution with a
special function ${\cal R}(t)$ that is selected by the short-time
difference effect of the time-derivatives. However, as time increases
and time-differences increase too the effect of the time-derivatives
diminishes. In the {\it approximate} analytic solution we take
advantage of this fact to solve the equations asymptotically but we
introduce in this way a symmetry that does not allow us to fix ${\cal
R}(t)$.  We obtain, instead, a family of solutions parametrized by
$h_t$.  It is important to reckon that the parametric constructions in
Fig.~\ref{fig:sketch-fig2} are independent of the clock and thus are
fully determined by the approximate treatment.

The case $p=2$ with Ising spins or Sherrington-Kirkpatrick model has a
more complicated scaling form with a sequence of two-time scales
leading to dynamic ultrametricity for all $C<q_{ea}$
asymptotically~\cite{Cuku2}. This
behaviour is technically more involved but, as far as the symmetry
properties are concerned, it is similar to the case treated above. The
full dynamic equations have a unique solution but the approximate ones
acquire time-reparametrization invariance. We shall not discuss these
cases further in the rest of this review.

\subsection{Short-ranged models -- dynamic action}

In a series of papers~\cite{Chamonetal1}-\cite{Jaubertetal} we claimed
that the time-reparametrization invariance thus far introduced {\it
via} the asymptotic solution to the dynamic equations in mean-field
glassy problems is indeed an asymptotic property of the dynamics of
glassy systems, mean-field {\it and} finite dimensional.  The
separation of time-scales stationary-aging has been observed in a
variety of glassy systems with numerical simulations and
experiments~\cite{aging,exps}. The slowness of the decay in the aging regime,
eqs.~(\ref{eq:slowness-C}) and (\ref{eq:slowness-chi})
and a {\it weak long-term memory} of the kind (\ref{eq:scaling-c}), are 
the hallmark of glassy relaxation. The idea is then that time-reparametrization
invariance {\it is} the symmetry associated to the dominant dynamic
fluctuations in these sytems.

In order to pursue this idea forward one has to first prove that the
symmetry of the saddle-point equations is also a symmetry of the
action in the dynamic generating functional not only of
fully-connected spin models of the mean-field type but also of finite
dimensional glassy systems. In~\cite{Chamonetal1} we derived and
studied the symmetry properties of the dynamic action -- the so-called
Martin-Siggia-Rose action associated to Langevin stochastic dynamics
-- of the disorder averaged soft-spin $3d$ Edwards-Anderson ({\sc ea})
model of spin-glasses. ($H=\sum_{\langle ij\rangle} J_{ij}s_i s_j$ 
with $J_{ij}$ Gaussian random variables with zero mean and taking 
non-zero value only on nearest neighbours on the $d$-dimensional lattice
and $s_i=\pm 1$.) In our analysis we took a number of steps
that we briefly recall here. 
First, we introduced four fluctuating two-time fields defined on the 
lattice sites,  
\begin{equation}
Q^{ab}_i(t,t_w) \; , \qquad \mbox{with} \qquad i=1,\dots, N \;\; 
\mbox{and} \;\; a,b=0,\, 1
\; .   
\end{equation} 
Their thermal averages are the expected values of the local two-time
self-correlation ($a=b=0$), the retarded linear response ($a=0$,
$b=1$), the advanced linear response ($a=1$, $b=0$), and a fourth
observable ($a=b=1$) that vanishes if causality is preserved.  Second,
we assumed that a separation of time-scales fast-slow, of the type
described in the previous subsection, applies to these fluctuating
fields too.  Third, we determined the long-time action by using a
Renormalization Group (RG) scheme in the {\em time} variables.  This
allowed us to write the full action as a sum of two contributions: one
from the fast regime holding at short time-differences, another one
from the slow regime valid at long time-differences. The coupling
between these two vanishes asymptotically.  Fourth, we analyzed the
surviving terms in the action, that are just the slow contribution. 
Using advanced and retarded scaling dimensions that are just the labels
$a$ and $b$ of the fields, one finds that
the {\it global time-reparametrization},
\begin{eqnarray}
t \to h_t \equiv h(t)
\; , \qquad\qquad
{{\widetilde{Q}}_i^{ab}}(t,t_w) = (\dot h_t)^a (\dot h_{t_w})^b \; 
Q_i^{ab}(h_t,h_{t_w}) 
\; , 
\label{eq:rpg1}
\end{eqnarray}
leaves all surviving terms unmodified. This step is concisely carried out 
as follows. Take a generic term in the action. Under (\ref{eq:rpg1})
it transforms as  
\begin{eqnarray*}
\int \prod_{\nu=1}^{\cal N} dt_\nu \; \cdots & \to & 
\int \prod_{\nu=1}^{\cal N} dt_\nu
\; 
(\dot h_{t_\nu})^{\Delta_\nu} \; \cdots
\end{eqnarray*}
where ${\cal N}$ is the number of time integrals, the dots 
represent a product of the fields $Q_i^{ab}$ and 
$\Delta_\nu$ is the sum of all factors $\dot h_{t_\nu}$
arising from the transformation of the fields. 
Interestingly enough, one finds that {\it all} the 
$\Delta_\nu$ equal one. Thus, with a simple change of 
variables one absorbs each factor $\dot h_{t_\nu}$ 
in the corresponding integration variable and 
\begin{equation}
\int \prod_{\nu=1}^{\cal N} dt_\nu \; \cdots \to  
\int \prod_{\nu=1}^{\cal N} dh_\nu \; \cdots 
\end{equation}
Note that in order to prove invariance under the 
simpler and more common scale transformation, $h_t=\lambda t$, it is 
enough to have 
$\sum_{\nu=1}^{\cal N} \Delta_\nu={\cal N}$. Scale invariance is 
included in the larger global time-reparametrization symmetry 
but it is, clearly, more restrictive. 
Finally, we showed that the measure in the
functional integral is also global time-reparametrization invariant, completing
the program. We refer the reader to Refs.~\cite{Chamonetal1} and 
\cite{Castilloetal2} for the
technical details leading to these results.

In the disordered 3$d$ {\sc ea} model studied 
we carried out the disorder average. The
presence of quenched disorder gave us an analytic control of the
theory but this does not necessarily mean that such a symmetry
develops only for the long time regime of models with quenched
disorder.  It appears that if the essential assumptions are causality
and unitarity, and a separation of time scales that takes the action
to a non-trivial asymptotic state (some glassy state), then one
expects the symmetry to exist for systems without quenched disorder
but that are glassy nevertheless. In order to check the development of
time-reparametrization invariance in problems of particles in
interaction in finite dimensions one should first obtain the relevant
action to work with. A good candidate for a starting point is the
Dean-Kawasaki stochastic equation for the evolution of the local
density~\cite{Dean}. The idea is then to write its dynamic generating
functional and study the symmetry properties of the effective action
assuming that a separation of time-scales exists.  We are currently
carrying out this study.

\subsection{Turning a nuisance into something useful - 
symmetry as a guideline}
\label{sym-useful}

The global time-reparametrization invariance implies that the action
describing the long time slow dynamics of a spin-glass is basically a
``geometric'' random surface theory, with the $Q$'s
themselves as the natural coordinates. The original two times
parametrize the surface. Physical quantities, as the bulk integrated
response $\chi(t_1,t_2)=\int_{t_2}^{t_1} dt' \, R(t',t_2)$ and
correlation $C(t_1,t_2)$ have scaling dimension zero under $t\to h(t)$
 as well as their local counterparts. 
The emergence of this gauge-like
symmetry, which may appear first as a nuisance that relates too many
solutions for just one problem. However, it may provide a simple way
to understand spatial fluctuations in systems that possess this {\it
global} time-reparametrization symmetry.

Here, a simple analogy with the problem of a ferromagnet may elucidate
the point we want to make. In a ferromagnet the action is invariant
under uniform rotations of the magnetization vector $\vec m$. In the
ordered phase, rotation symmetry is spontaneously broken, and a
certain magnetization direction $\vec m_0$ is picked. Typically, a
vanishingly small pinning field selects this direction.  The low
action excitations are the spin waves, fluctuations of the uniform,
symmetry broken, state. These spin waves can be described in terms of
smooth spatially fluctuating rotations around the uniform magnetization
state. The spin waves, generated by using slowly varying local
rotations, are the Goldstone modes of the ferromagnet.

Similarly, in the aging regime of the glassy systems, the action has a
global symmetry, under uniform time-reparametrizations, $t\to h(t)$,
with the fields transforming as in eq.~(\ref{eq:rpg1}). The probability
weight of having certain {\it local} two-time correlation and
response, the observables, should be independent of this
reparametrization. After coarse-graining over a 
linear length $\ell$  the non-vanishing fluctuating fields are
the local {\it coarse-grained} correlation ($a=b=0$) 
and linear response ($a=0,\, b=1$),
\begin{displaymath}
C(\vec r;t,t_w) = \left( \frac{a}{\ell}\right)^d 
\sum_{j\in V_{\vec r}} s_j(t) s_j(t_w) \; , \;\;\; 
R(\vec r;t,t_w) = \left( \frac{a}{\ell}\right)^d 
\sum_{j\in V_{\vec r}} \left. \frac{\delta s_j(t)}{\delta
h_j(t_w)}\right|_{h=0} \; ,
\end{displaymath}
with the sum carried over the spins in the volume $V_{\vec r}=\ell^d$ 
centered at
$\vec r$, and $a$ is the lattice spacing. The transformation (\ref{eq:rpg})
is now restated as 
\begin{eqnarray}
t \to h_t \equiv h(t)
\; , \qquad
&&
\left\{ 
\begin{array}{l} 
C_{ag}(\vec r;t,t_w) \to C_{ag}(\vec r;h_t,h_{t_w}) \; , \\
R_{ag}(\vec r;t,t_w) \to \dot h_{t_w} \, R_{ag}(\vec r;h_t,h_{t_w}) \; 
\end{array},
\right.
\end{eqnarray}
and it is an {\it asymptotic} symmetry of the action for the slow
coarse-grained degrees of freedom. Indeed, the
symmetry breaking terms, that have their origin in the short-time
dynamics and short-time difference dynamics, are not identical to zero
but become vanishing small asymptotically. The particular scaling
function ${\cal R}(t)$ selected by the system is determined by
matching the fast and the slow dynamics. It depends on several details
-- the existence of external forcing, the nature of the microscopic
interactions, {\it etc.} In other words, the fast modes which are
absent in the slow dynamics act as symmetry breaking fields for the
slow modes.

In analogy with the spin-wave fluctuations in magnetic systems, that are
dictated by the rotational symmetry, we proposed that the smooth
fluctuations in the glassy phase can be obtained by studying the slow
varying, position dependent reparametrizations $t\to h(\vec{r},t)$
around the one reparametrization ${\cal R}(t)$ selected by the
short-time dynamics. In other words, we basically proposed that there
are Goldstone modes for the glassy action which can be written as
slowly varying, spatially inhomogeneous time reparametrizations.
This suggests that the slow
part of the coarse-grained {\it local} correlations and
susceptibilities should scale as
\begin{eqnarray}
&&
C_{ag}(\vec r;t,t_w)  \approx
q_{ea} \; f_{\sc c}\left(\frac{h(\vec r,t)}{h(\vec r,t_w)}\right)
\; , \qquad 
{\chi_{ag}}(\vec r;t,t_w) \sim
f_\chi\left( \frac{h(\vec r,t)}{h(\vec r,t_w)}\right)
\; ,
\label{eq:fluct-cchi}
\end{eqnarray}
with $f_c$ and $f_\chi$ the {\it same} functions describing the global
correlation and susceptibility, respectively [eqs.~(\ref{eq:scaling-c})
and (\ref{eq:chiag})] and the same function $h(\vec r,t)$ scaling the
two-time correlation and susceptibility on each site
$\vec r$~\cite{Chamonetal1,Castilloetal1,Castilloetal2}.  The sum rules
$C_{ag}(t,t_w) = V^{-1} \int d^dr \; C_{ag}(r;t,t_w)$ and
$\chi_{ag}(t,t_w) = V^{-1} \int d^dr \; \chi_{ag}(r;t,t_w)$ apply.

The reason for this proposal is that the {\it global}
reparametrization invariance in time of the dynamic action in this
two-time regime leads to low action excitations (Goldstone modes) for
smoothly varying {\it spatial} fluctuations in the reparametrization
of time, but not in the external form of the scaling functions. As in
a sigma model (for example, to describe the ferromagnet), the external
functions $f_c$ and $f_\chi$ fix the manifold of states, and the local
time reparametrizations correspond to fluctuations restricted to this
fixed manifold of states (in the ferromagnet, the tilting of direction
but not the magnitude of the magnetization vector).

\subsection{The spherical $p=2$ case or mean-field domain growth}
\label{subsec:p=2}

The spherical SK model ($p=2$) 
can be solved {\it exactly} by analyzing the Langevin
equation in the basis of eigenvectors of the random matrix
$J_{ij}$~\cite{Cude}.  While the correlation has a very similar
behaviour to the one of the $p\geq 3$ cases (see
Fig.~\ref{fig:sketch-fig}-left), the susceptibility is quite
different. One can mention that the stationary and aging regimes 
in the linear response are
not so sharply separated in this case. If one uses an additive
separation as in (\ref{eq:separation}) the aging contribution to the
integrated linear response, $\chi_{ag}(t,t_w)$, vanishes
asymptotically. More precisely,
\begin{equation}
\chi_{ag}(t,t_w) \sim t_w^{-1/2} \; f_\chi\left( \frac{t}{t_w}\right)  
\; .
\label{eq:chiag}
\end{equation}

Importantly enough, even though the inequalities (\ref{eq:slowness-C})
and (\ref{eq:slowness-chi}) are still valid, a careful inspection of
all terms in the Schwinger-Dyson equations shows that they are of the
same order asymptotically. Moreover, the stationary contribution to
the equations in the aging regime is not just a constant: the
corrections associated to the asymptotic approach to the plateau in
the correlation cannot be neglected. As a result one cannot simply
drop the time derivatives and  the Schwinger-Dyson equations
in the aging regime {\it are not} time-reparametrization invariant but
just {\it scale invariant}, that is to say, they are unchanged by the
transformation $t\to h(t)=\lambda t$, with $C$ and $R$ transforming as in
(\ref{eq:rpg}) and $\lambda$ a positive constant.
A similar mechanism, though even harder to prove, applies to the
dynamic equations for the correlation and linear response of the
non-conserved dynamics of the O($N$) ferromagnetic model in the large
$N$ limit~\cite{Chamonetal3}.

In line with what we explained above, the effective action for the
slow degrees of freedom of the $p=2$ spherical model and, more
generally, the $d$-dimensional ferromagnetic $O(N)$ model in the large
$N$ limit, are not invariant under global time-reparametrizations but
only under global rescaling of time, $t\to \lambda t$. This marks an
important difference between models with a finite aging response and
these quasi quadratic models with a vanishing aging
response~\cite{Chamonetal3}.

This result is important for a number of reasons: first, it suggests
that the susceptibility, or even the effective temperature, might be
intimately related to the symmetry properties of the dynamics and
consequently of the fluctuations; second, it suggests that the
mechanism for fluctuations in coarsening systems might be different
from the one of other glassy problems with finite and well-defined
effective temperatures. In order to justify the latter statement it
remains to be checked whether the reduction of time-reparametrization
invariance to just scale invariance also holds in finite dimensional
non-field coarsening. 

\subsection{Quantum problems}

In the case of {\it quantum} models one introduces the effect of
dissipation by coupling the system to an environment represented,
typically, by an infinite ensemble of quantum harmonic
oscillators. One then uses the Schwinger-Keldysh formalism to write a
generating functional and from it, in the fully-connected or infinite
dimensional cases, one derives Schwinger-Dyson equations similar to
the ones above. The asymptotic analysis of these equations follows the
same steps as in the classical limit~\cite{Culo,Chke} (at least in the
case of a weak coupling to the bath~\cite{quantum-bath}) and the
time-reparametrization invariance also applies.

The appearance of an asymptotic invariance under time-reparatrizations
in the mean-field dynamic equations was related to the
reparametrization invariance of the replica treatment of the statics
of the same models~\cite{Sompo,Brde}.  The latter remains rather
abstract. Br\'ezin and de Dominicis~\cite{Brde} studied the
consequences of twisting the reparametrizations in the replica
approach. Interestingly enough, this can be simply done in a dynamic
treatment either by applying shear forces
or by applying heat-baths with different
inherent dynamics to different
parts of the system. More precisely, using a model with open boundary
conditions one could apply a thermal bath with a characteristic
time-scale on one end and a different thermal bath with a different
characteristic time-scale on the opposite end and see how a
time-reparametrization `flow' establishes in the model.

\section{Consequences and tests} 
\label{sec:consequences}

In this Section we discuss how one can put these ideas to the test by
presenting a number of consequences of global time-reparametrization
invariance that are directly measurable numerically and
experimentally. The properties that we discuss explicitly are:
\begin{enumerate}

\item[{\it 1.}] A growing dynamical correlation length.

\item[{\it 2.}] Scaling of the {\sc pdf} of local two-time functions.

\item[{\it 3.}] Functional form of the {\sc pdf} of local two-time functions.

\item[{\it 4.}] Triangular relations between two-time functions.

\item[{\it 5.}] Scaling relations for general multi-time functions.

\item[{\it 6.}] Local fluctuation-dissipation relations.

\item[{\it 7.}] Infinite susceptibilities.

\end{enumerate}

In considering these predictions, we shall separate them into two
distinct classes. The first class contains predictions that are
consistent with other theoretical scenarios {\it as well as} with the
presence of reparametrization invariance. Hence, while
reparametrization invariance leads to these predictions, this class
alone cannot be used to argue for the role of the symmetry over other
mechanisms. The second class, on the other hand, contains predictions
that are not natural within other frameworks, and to the date of this
report have no obvious explanation within other frameworks. Properties
{\it 1} and {\it 2} belong to the first class, properties {\it 3}-{\it
7} to the second. We shall present these properties in detail below.


\subsection{Two-time correlation length}
\label{eq:two-time-length}

In equilibrium statistical models one defines the {\it static
correlation length}, $\xi_{eq}$, from the spatial decay of the
correlation between the fluctuations of the order parameter measured
at two space points, $\langle \, [\phi(\vec r) -\langle \, \phi(\vec
r) \, \rangle] [\phi(\vec{r'}) -\langle \, \phi(\vec{r'}) \, \rangle]
\, \rangle_{|\vec r-\vec{r'}|=\Delta} \sim \Delta^{-d+2-\eta} \;
e^{-\Delta/\xi_{eq}}$, where the angular brackets denote an average
over the Gibbs-Boltzmann measure.  $\xi_{eq}$ depends on temperature
and, in second order phase transitions, it diverges at $T_c$ leaving
only an algebraically decaying correlation. 
A {\it dynamic equilibrium correlation length} characterizes the
spatial decay of equal-time correlations in the equilibrium relaxation
of `usual' systems. Similarly to the static case one defines $\xi(t)$
{\it via} $\langle \, [\phi(\vec r,t) -\langle \, \phi(\vec r,t) \,
\rangle] [\phi(\vec{r'},t) -\langle \, \phi(\vec{r'},t) \, \rangle] \,
\rangle_{|\vec r-\vec{r'}|=\Delta} \sim e^{-\Delta/\xi(t)}$; the
angular brackets indicate here an average over thermal histories and,
for simplicity, we omitted the algebraic correction to the exponential
decay.  The average over thermal noise can be traded for an
integration over the reference space-point $\vec r$ and one then 
obtains $\xi(t)$ from $V^{-1} \int d^d r \; \delta \phi(\vec r,t)
\delta \phi(\vec{r'},t)\; |_{|\vec r-\vec{r'}|=\Delta} \sim
e^{-\Delta/\xi(t)}$ where $\delta \phi(\vec r,t) \equiv \phi(\vec r,t)
- V^{-1} \int d^dr'' \; \phi(\vec{r''},t)$.  The correlation length,
$\xi(t)$, depends now on temperature and total time.

In systems with slow dynamics in which the order parameter is a
{\it two-time} entity, a two-time correlation length can be defined 
in analogy to what we described in the previous paragraph:
\begin{eqnarray}
&&
\left\langle
\vphantom{\phi(\vec{r'})}
\; 
[\phi(\vec r, t)\phi(\vec r, t_w)
- \langle \, \phi(\vec r, t)\phi(\vec r, t_w)
\, \rangle ]
\right.
\nonumber\\
&&
\qquad 
\times
\left. 
[
\phi(\vec{r'}, t)\phi(\vec{r'}, t_w)
- \langle \, \phi(\vec{r'}, t)\phi(\vec{r'}, t_w) 
\, \rangle 
]
\; 
\right\rangle_{|\vec r-\vec{r'}|=\Delta} 
\sim e^{-\Delta/\xi(t,t_w)}
\; .  
\end{eqnarray}
Once again by trading the thermal average by a spatial 
average  $\langle \, \phi(\vec r, t)\phi(\vec r, t_w) \, \rangle$
becomes the global correlation $C(t,t_w)$ and 
$\xi(t,t_w)$ is derived from the four-point correlation
\begin{eqnarray}
&&
C_4(\Delta; t,t_w) \equiv
\frac{1}{V}
\int d^dr
\, 
\left.
\delta [\phi(\vec r, t)\phi(\vec r, t_w)]
\delta [\phi(\vec{r'}, t)\phi(\vec{r'}, t_w)]
\;
\right|_{|\vec r-\vec{r'}|=\Delta} 
\end{eqnarray}
with $\delta[\phi(\vec r, t)\phi(\vec r, t_w)] 
\equiv \phi(\vec r, t)\phi(\vec r, t_w) - C(t,t_w)$. The quantity
$C_4(\Delta; t,t_w)$ measures the probability that a fluctuation of
the two-time composite field $\phi(\vec r, t)\phi(\vec r, t_w)$ 
with respect to its global average
$C(t,t_w)$ in the spatial position $\vec r$ 
affects a fluctuation of the same composite field at a
different site $\vec{r'}$ located at a distance $\Delta$ from $\vec r$
and averaged over the whole ensemble of reference points $\vec r$ in
the sample.  

\begin{figure} 
\centerline{ 
\includegraphics[width=5cm]{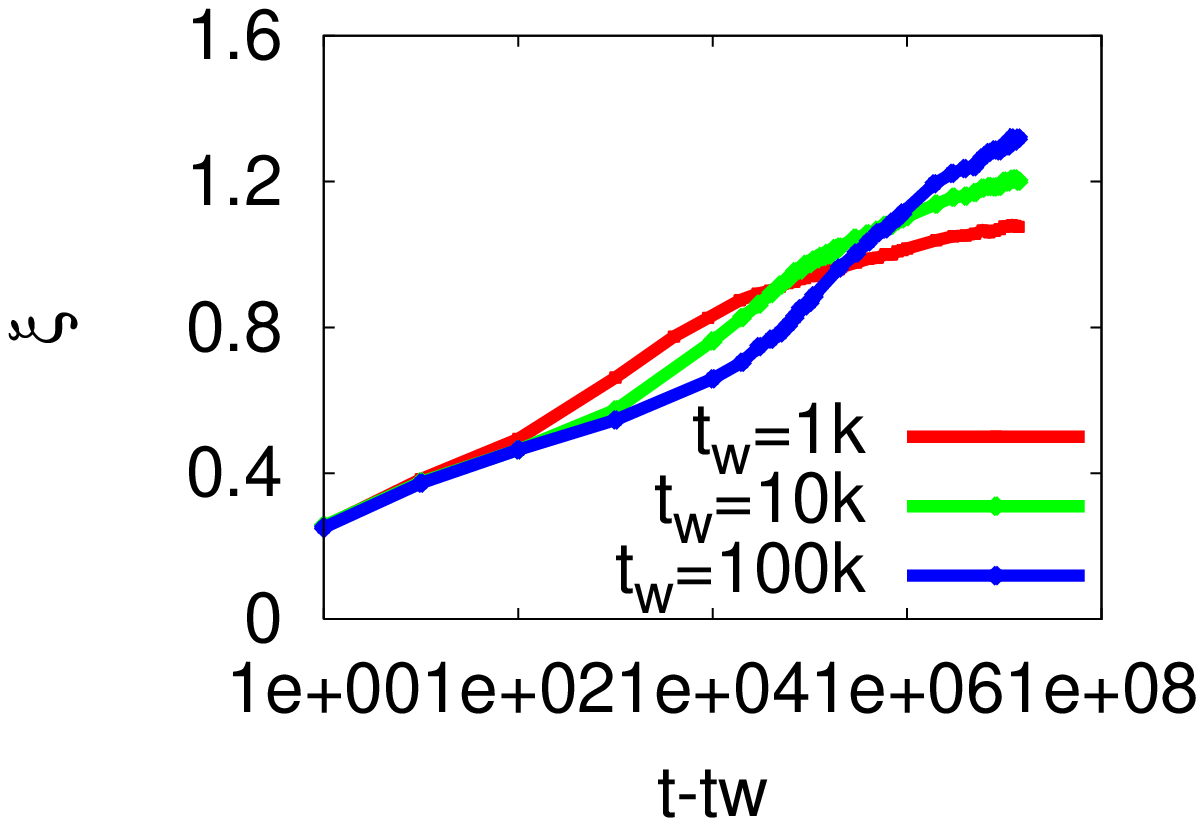} 
\includegraphics[width=5cm]{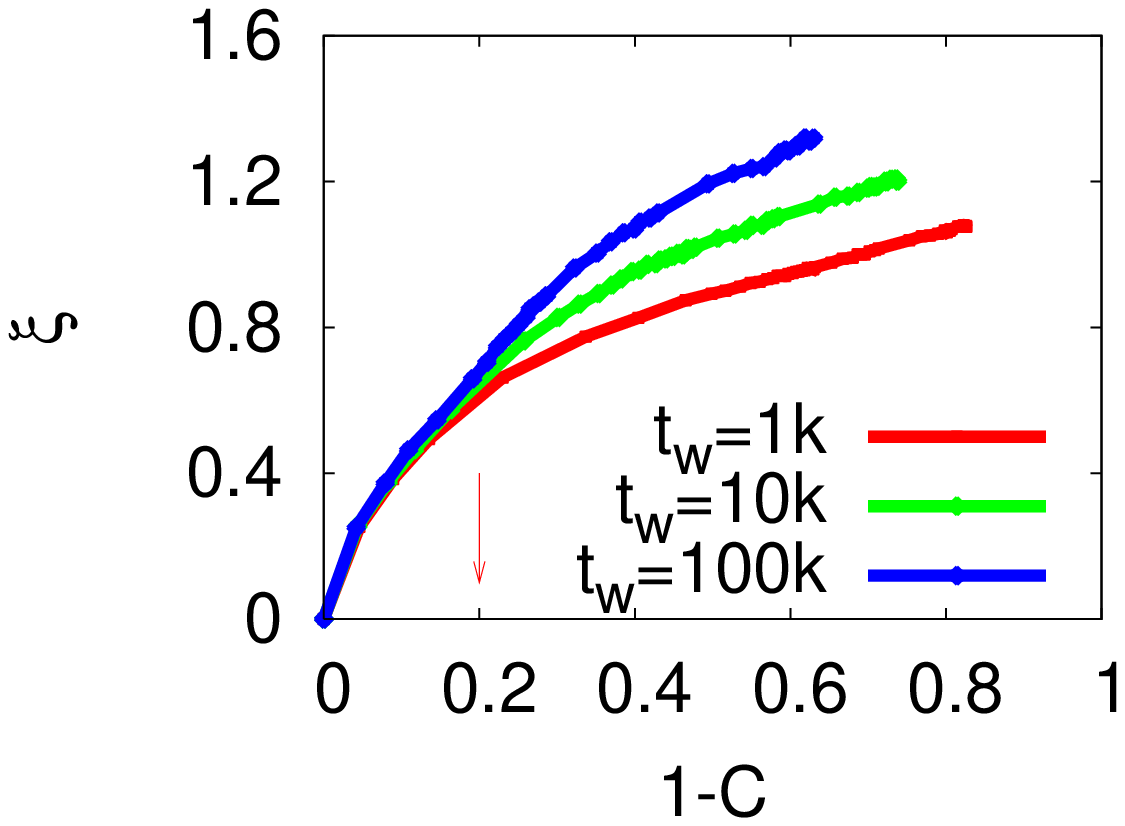} 
\includegraphics[width=5cm]{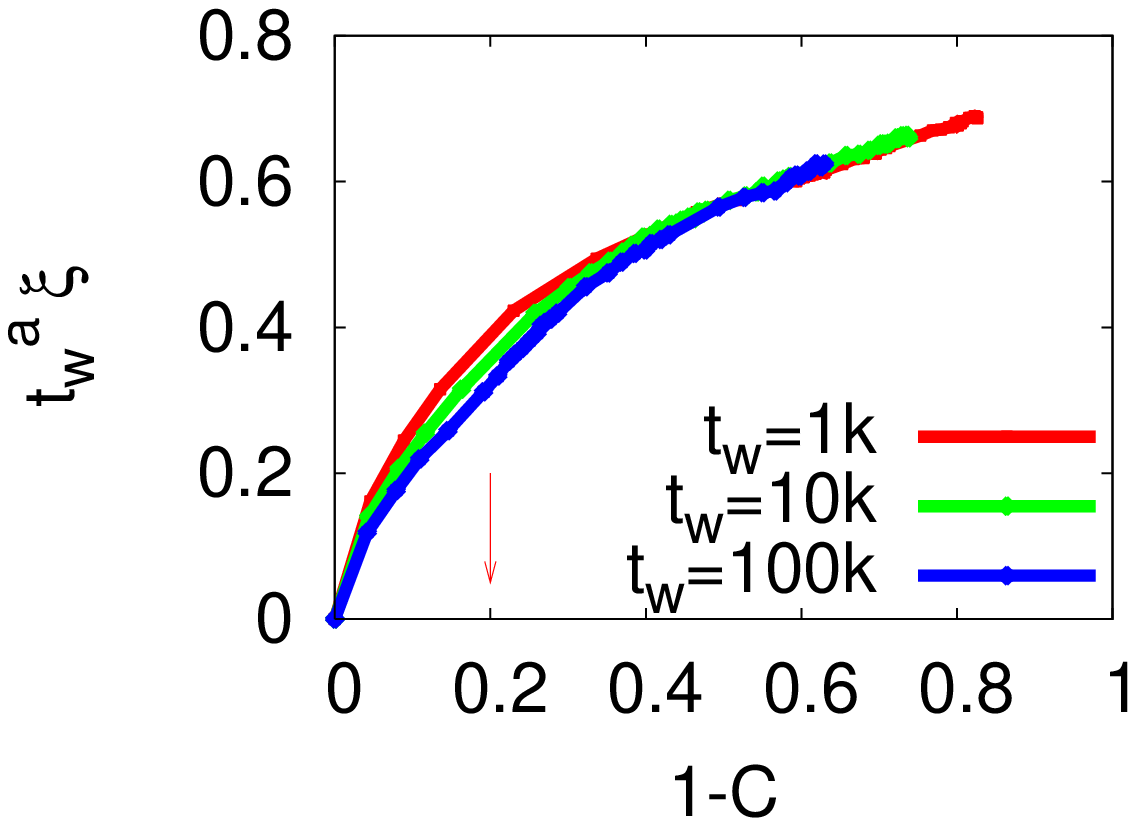} 
} 
\caption{The correlation length in the $3d$ EA model
at $T=0.6<T_c$ and $L=100$. 
(a) As a function of $t-t_w$; (b) as a function of $1-C$; 
(c) in the scaling form $t_w^{-a} \xi$ against 
$1-C$. These results are taken from~\cite{Jaubertetal}. The 
vertical arrow in panels (b) and (c) indicates the value of $q_{ea}$.}
\label{fig:longueur} 
\end{figure}

The numerical analysis of $C_4$ in 
the low temperature out of equilibrium dynamics 
of the $3d$ {\sc{ea}} model~\cite{Castilloetal2,Jaubertetal} 
(see Fig.~\ref{fig:longueur}), 
soft sphere~\cite{Parisi} 
and Lennard-Jones~\cite{Castillo2} mixtures yield
\begin{eqnarray} 
\xi(t,t_w) &\sim& 
\left\{
\begin{array}{ll}
\xi_{st}(t-t_w) \; \qquad &C>q_{ea}
\nonumber\\
\xi_{ag}(t,t_w) \; \qquad &C<q_{ea}
\end{array}
\right.
\qquad \mbox{with}
\nonumber\\
\xi_{ag}(t,t_w) 
&\sim& t_w^a \; g(C) \;\; \mbox{and} \;\; a \;\;\mbox{a small power}
\end{eqnarray}
(a logarithmic growth is also possible).  
$g(C)$ is a monotonically decreasing function of $C$. This is a two-time
monotonically growing function even at time-lags that are longer than
the waiting-time dependent $\alpha$ relaxation time. Note the
difference with what is observed in the super-cooled phase where a
number of numerical and experimental measurements point at a dynamic
correlation length that reaches a maximum at the $\alpha$ relaxation
time and later diminishes to zero~\cite{Lacevic}.  The divergence has
a clear interpretation within the global time-reparametrization
scenario: it is due to the generation and development of the zero
mode. The global reparametrization invariance symmetry develops only
in the limit of very long times, so that at intermediate times the
irrelevant terms that are scale down to zero are still
manifest. These irrelevant terms, symmetry breaking ones, give a
finite length scale (or a finite `mass') to the soft reparametrization
modes. Because they are irrelevant, the correlation length increases
or, equivalently, the mass decreases asymptotically.

Even though the numerically accessible times are sufficiently long so
as to see the separation of time-scales in the relaxation of the
global correlation, the correlation length is still very short. In
numerical simulations $\xi$ reaches of the order of 4 lattice spacings
or inter-particle distances in the spin-glass problem or soft sphere
and Lennard-Jones system, respectively.  $\xi$ just increases by, say,
a factor $4$ when the waiting-time is increased by nearly 4 orders of
magnitude.  The expected limit $\xi(t,t_w) \to\infty$ is thus far from
being attained.


As we already mentioned in the introduction to this Section, the
growing length scale is consistent with the development of
time-reparametrization invariance, but it is {\it also} consistent
with other mechanisms. Hence, it alone cannot be used to argue
unambiguously in favor of the symmetry-based approach. For example,
theories based on the mode-coupling approach or random first order
scenario~\cite{Silvio-length} and its refinement including the effect
of entropic droplets~\cite{Wolynes}, dynamical criticality controlled
by a zero-temperature critical point~\cite{Garrahan-etal} and frustration
limited domains~\cite{Tarjus-etal} are used to justify the growth of a
dynamic length-scale, at least in the super-cooled liquid. Hence, the
existence of a growing length scale belongs to the first class of
predictions we mentioned in the introduction to this section.

Finally, let us note that the fluctuations in the susceptibility,
or in  multi-time correlations, see eq.~(\ref{multi-time}), and 
associated susceptibilities, can be used to derive other
correlation lengths.  It would be interesting to check whether all
these behave in the same manner.

\subsection{Scaling of the {\sc pdf} of local two-time functions}

The most direct way of testing the mere existence of local
fluctuations is to measure the probability distribution function ({\sc
pdf}) of local coarse-grained correlators, $C(\vec r;t,t_w)$, and linear
susceptibilities, $\chi(\vec r;t,t_w)$, at different pairs of times, $t$
and $t_w$.  In such a measurement one is forced to use finite
coarse-graining lengths. $\ell$ then becomes a parameters that has to
be taken into account in the scaling analysis of the results.

In Ref.~\cite{Jaubertetal} we showed that, quite generally, the {\sc pdf}
of local coarse-grained correlators can be scaled onto universal
curves as long as the global correlation, $C(t,t_w)$, is the same, and
the ratio of the coarse graining length over the dynamical correlation
length, $\ell/\xi(t,t_w)$, is held fixed (see~\cite{Berthier} for a similar 
discussion applied to the super-cooled liquid).  Such scaling can be
easily understood as follows. At fixed temperature the {\sc pdf}
$\rho[C_r;t,t_w,\ell,L]$ depends on four parameters: two times, $t$
and $t_w$, and two lengths, the coarse-graining length, $\ell$, and 
the size of the system, $L$.  As in the aging regime 
$C(t,t_w)$ is a monotonic function of the two times and $\xi(t,t_w)
\sim t_w^a g(C)$, one can trade the two times by
$C$ and $\xi$ in complete generality.  The next step is a {\it scaling
assumption}: that in the long times limit the {\sc pdf}s depend on the
coarse graining length $\ell$, the total size $L$ and the scale $\xi$
only through the ratios $\ell/\xi$ and $\xi/L$. This last step, we
should stress, is really a scaling assumption, and not a trivial
requirement from dimensional analysis. The lengths $\ell$, $L$ and
$\xi$ are already dimensionless as they are measured in units of the
lattice spacing.  The end result from the rewriting of the parameters
in terms of the global correlation and the scaling hypothesis is that
the {\sc pdf}s characterizing the heterogeneous constant temperature
aging of the system can be written as
\begin{equation} 
\rho[C_r;C(t,t_w), \ell/\xi(t,t_w),\xi(t,t_w)/L] 
\; .
\label{eq:pdf-scaling1}
\end{equation}

\begin{figure} 
\centerline{ 
\includegraphics[width=7cm]{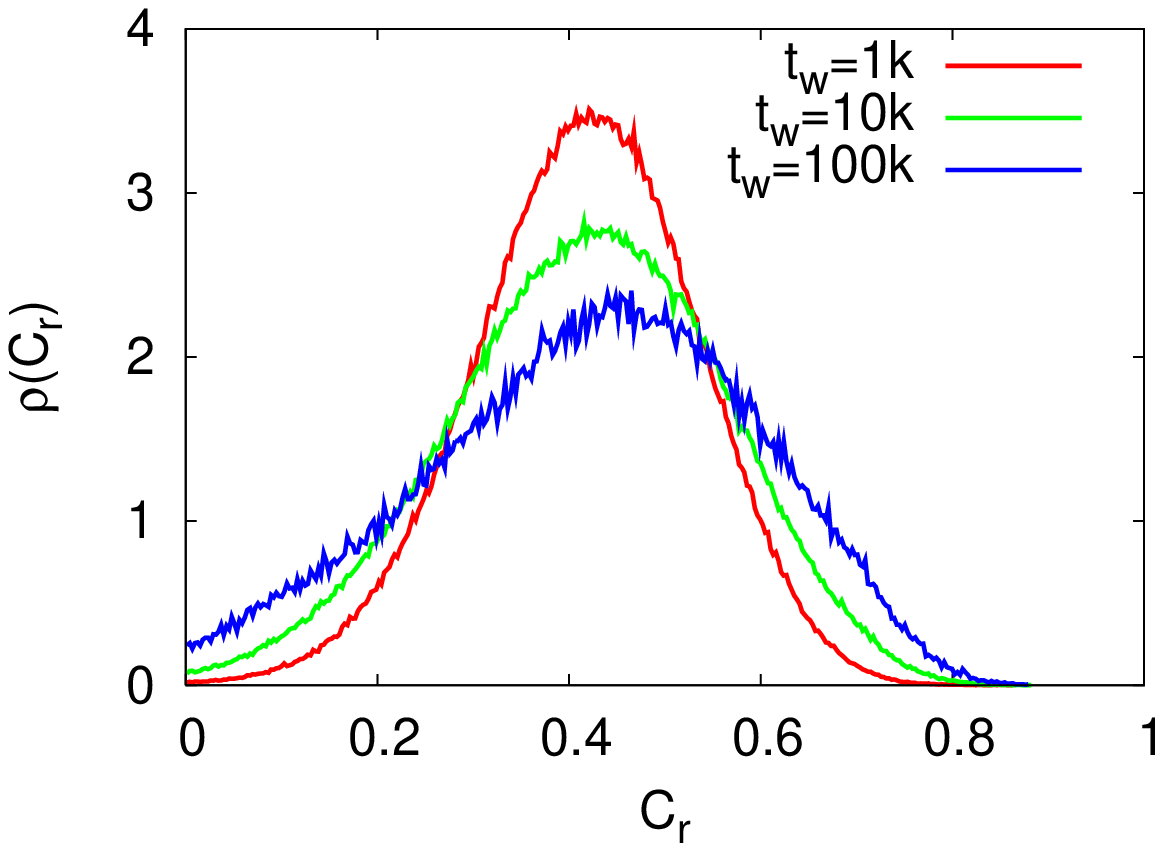} 
\includegraphics[width=7cm]{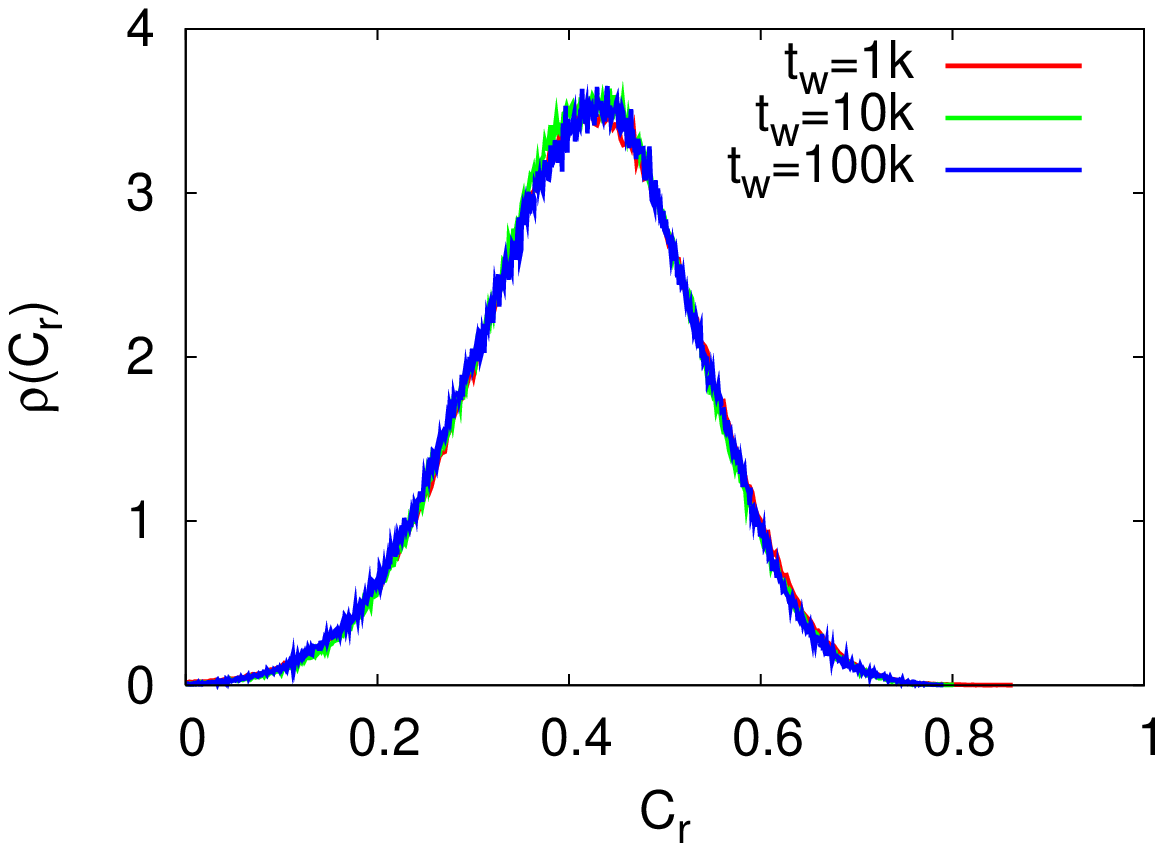} 
} 
\caption{{\sc pdf} of local coarse-grained correlations $C_r$ at
different times $t$ and $t_w$ in the $3d$ Edwards-Anderson model
with
 $L=100$ at $T=0.6<T_c$.  
The waiting-times are given in the key
 and the
global correlation is fixed to $C=0.4<q_{ea}$.  (a) The
coarse-graining boxes have linear size $\ell=9$ in all cases.  The
curves do not collapse, a slow drift with increasing $t_w$ is clear
in
 the figure.  (b) Variable coarse-graining length $\ell$ chosen so
as
 to held $\ell/\xi$ approximately constant.  The collapse improves
considerably with respect to panel
 (a). These results are taken
from~\cite{Jaubertetal}.}
\label{fig:time-scaling-pdfs} 
\end{figure}

In Fig.~\ref{fig:time-scaling-pdfs} we show the scaling of the distribution of 
local coarse-grained
correlations in the $3d$ {\sc ea} model and the effect of the
scaling variable $\ell/\xi$ (the size of the system, $L$, is sufficiently 
large so that $\xi/L$ vanishes in practice). 
 It is noteworthy that a reasonable scaling with the global
correlation held fixed and not taking into account the effect of the
second scaling variable has been already achieved, approximately, in
the Edwards-Anderson model~\cite{Castilloetal1,Castilloetal2}, as well
as in the kinetically constrained models studied in \cite{Chamonetal2}
and Lennard-Jones systems~\cite{Castillo1}. This is justified by the
fact that $\xi$ varies very slowly with $t_w$.  However, it is clear that
at long though finite times one has to hold the ratio $\ell/\xi$
constant to obtain a full collapse in all cases.

The scaling variable $\ell/\xi$ allows one to study the change 
in the functional 
form of the {\sc {pdf}}s upon modifying the
coarse-graining volume. Indeed, the {\sc pdf}s
should crossover from a
non-trivial form to a simple Gaussian when $\ell$ goes through the
value $\xi$.
In summary, for {\it finite} $\xi$ one identifies three 
$\ell$-dependent regimes with different functional forms
of the {\sc pdf}s:
\begin{itemize} 
\item 
$\ell \ll \xi$.
For finite $\xi$ this means $\ell$ of the order of the 
lattice spacing, $\ell\sim a$. In this case the {\sc pdf}s do not 
have any particular structure. 
\item 
$\ell \sim \xi$. For finite but large $\xi$ this case is non-trivial
and indeed the one that is accessed with numerical simulations and
experiments. One finds that the statistics is non-Gaussian for all $C$.
The skewness decreases from zero at $C=1$ to reach a minimum and 
then increase again at small values of $C$.
As regards the functional form, one finds that a
Gumbel-like functional form, characterized by a real parameter that
depends on $C$ and $\ell/\xi$ describes the data rather well for, say,
$q_{ea}/2 \stackrel{<}{\sim} C \stackrel{<}{\sim} q_{ea}$.

\item 
$\ell \gg \xi$. In this limit one matches the central-limit theorem
conditions and the statistics becomes Gaussian for all $C$. 
\end{itemize}
The analysis of the $\ell$-dependent {\sc pdf}s thus
provides an independent way to estimate $\xi$.


\subsection{Effective action for local ages}

The argument leading to eq.~(\ref{eq:pdf-scaling1}) is a scaling
hypothesis and it does not rely on time-reparametrization
invariance. Indeed, there exist models in which the
scaling~(\ref{eq:pdf-scaling}) for the {\sc pdf} of local
coarse-grained correlations is found, {\it e.g.} the $O(N)$ model in
the large $N$ limit~\cite{Chamonetal3}, 
and global time-reparametrization does not
apply. 

The implications of time-reparametrization
invariance appear later, as a 
prediction for the functional form of the asymptotic limit
\begin{equation}
\rho_\infty(C_r;C) \equiv
\lim_{\ell\to\infty}
\lim_{\stackrel{t,t_w\to\infty}{C(t,t_w)=C}} 
\lim_{L\to\infty} \; 
\rho[C_r;C(t,t_w), \ell/\xi(t,t_w),\xi(t,t_w)/L] 
\; . 
\label{eq:pdf-scaling}
\end{equation}

In order to study the functional form that $\rho_\infty$ can take let us
now use the symmetry argument to analyse the statistics of the
fluctuations of the local correlations. So far we have not yet
determined how much the $h(\vec r,t)$ vary in space and time. To this end
we need to derive an effective action for these functions that will
tell us how costly it is to deviate from the average clock ${\cal
R}(t)$. Ideally, one would like to derive this action from the
microscopic one. This should be possible in quasi mean-field models
such as the $p$-spin model with Kac long (but finite) range
interactions~\cite{Silvio}. For the moment we have just proposed the
simplest action that serves our purposes
in the ideal limit in which the zero
 mode is fully developed $\xi\to\infty$ and the local quantities are
 measured in the infinite coarse-graining volume limit with
 $\ell/\xi\to 0$~\cite{Chamonetal2}.  
Otherwise the parameter $\ell$ should be taken into
 account.

To start with we worked with
the more convenient transformed variable $h(\vec r,t) \equiv
e^{-\varphi(\vec r,t)}$ that implies
\begin{eqnarray}
&&
C_{ag}(r;t,t_w) \approx q_{ea} \; 
f_{\sc c}\left( \frac{h(\vec r,t)}{h(\vec r,t_w)} \right)
= 
q_{ea} \; f_{\sc c}
\left( e^{-\int_{t_w}^t dt' \partial_{t'} \; \varphi(\vec r,t')}  \right)
\label{eq:local-corr-param}
\end{eqnarray}
and we searched for the simplest action that satisfies the constraints
due to the symmetries. These are:
\begin{enumerate}

\item[{\it i.}] The action must be invariant under a global time
reparametrization $t\to h(t)$.

\item[{\it ii.}] If our interest is in short-ranged problems, the
action must be written using local terms. The action can thus contain
products evaluated at a single time and point in space of terms such
as $\varphi(\vec r,t)$, $\partial_t\varphi(\vec r,t)$, 
$\nabla\varphi(\vec r,t)$,
$\nabla\partial_t\varphi(\vec r,t)$, and similar derivatives.

\item[{\it iii.}] The scaling form in eq.~(\ref{eq:local-corr-param})
is invariant under $\varphi(\vec r,t)\to\varphi(\vec r,t)+\Phi(\vec
r)$, with $\Phi(\vec r)$ independent of time.  Thus, the action must
also have this symmetry.

\item[{\it iv.}] The action must be positive definite.

\end{enumerate}

These requirements largely restrict the possible actions. The one
with the smallest number of spatial derivatives (most relevant terms)
is
\begin{equation}
{\cal S}[\varphi]=
\int d^dr \int dt 
\left[
K\; \frac{\left(\nabla\partial_t\varphi(\vec r,t)\right)^2}
{\partial_t\varphi(\vec r,t)}
\right]\; ,
\label{eq:S-effective}
\end{equation}
with $K$ a stiffness. A term $M\; \partial_t\varphi(\vec r,t)$
is also allowed by symmetry but since its
space-time integral is constant we drop it. 
The action solely depends on the time derivatives
$\partial_t\varphi(\vec r,t)$ and it is simple to 
check that it satisfies all the four constraints enumerated above
(the last requirement follows from the fact that 
$h(\vec r,t)$ are monotonically increasing functions of
time)~\cite{Chamonetal2}.

Due to the simple form (\ref{eq:S-effective}) the $\partial_t
\varphi(\vec r,t)$ are uncorrelated at any two different times $t_1$
and $t_2$. Thus the expression $\Delta\varphi_{\vec
r}|_{t_w}^t\equiv\int_{t_w}^t dt' \; \partial_{t'} \varphi(\vec r,t')$
entering the exponential in the scaling form in
eq.~(\ref{eq:local-corr-param}) is a sum of uncorrelated random
variables in time. One can interpret such expression as the
displacement of a random walker with position dependent
velocities. Alternatively, one can think of the space-dependent
differences $\Delta\varphi_{\vec r}|_{t_w}^{t}$ as the net
space-dependent height (labeled by $t$) of a stack of spatially
fluctuating layers $dt\, \partial_t \varphi(\vec r,t)$. The action for
the fluctuating surfaces of each layer is given by
eq.~(\ref{eq:S-effective}).

The statistics of the $\Delta\varphi_{\vec r}|_{t_w}^t$ are completely
determined as follows. The action (\ref{eq:S-effective}) transforms
into one of a Gaussian surface after the introduction of a `proper'
time $\tau \equiv \ln {\cal R}(t)$, and the change of variables,
$\psi^2(\vec r,t)= \partial_\tau \varphi(\vec r,\tau)$. Indeed, 
\begin{eqnarray}
&& C_{ag}(\vec r;t,t_w) \approx f_{\sc c}
\left( e^{-\int_{\ln {\cal R}(t_w)}^{\ln {\cal R}(t)} d\tau' \; 
\psi^2(\vec r,\tau')} \right)
\label{eq:C-shape}
\\
&& {\cal S}[\psi] = 
K \int d^d r \int d\tau' \; [\nabla \psi(\vec r,\tau')]^2
\; . 
\label{eq:Action-shape}
\end{eqnarray}
Due to the Gaussian statistics of the $\psi$, it is simple to
show that connected $N$-point correlations of 
$\Delta\\varphi_{\vec r_1}|^t_{t_w}$
satisfy
\begin{eqnarray}
\langle
\Delta\varphi_{r_1}|^t_{t_w}\;
\Delta\varphi_{r_2}|^t_{t_w} \cdots 
\Delta\varphi_{r_N}|^t_{t_w}
\rangle_c
= 
&&[\tau(t)-\tau(t_w)]\; {\cal F}(\vec r_1,\vec r_2,\dots,\vec r_N)
\; ,
\label{N-corr}
\end{eqnarray}
where the function ${\cal F}$ can be obtained from Wick's theorem,
summing over all graphs that visit all sites (connected) with two
lines (because of $\psi^2$) for each vertex $i$ corresponding to a
position $r_i$. The reparametrized times appear only in the prefactor
$\tau(t)-\tau(t_w)=\ln [{\cal R}(t)/{\cal R}(t_w)]$.  The
probabilistic features of the fluctuations of local correlations
$C(\vec r,t,t_w)$ depend on times only through ${\cal R}(t)/{\cal R}(t_w)$,
and hence only through the global correlation itself $C(t,t_w)$.
In consequence, the action~(\ref{eq:Action-shape}) 
implies the scaling~(\ref{eq:pdf-scaling}).  The fact that the
time-dependencies of the statistical properties of the two-time local
coarse-grained correlations are fully determined by the global
correlation is a very welcome property of
action~(\ref{eq:S-effective}) since it was not obvious {\it a priori}.

Having the forms in eqs.~(\ref{eq:C-shape}) and
(\ref{eq:Action-shape}) allows us to make some quantitative
predictions about the form of the {\sc pdf}s. 
With some algebraic manipulations one shows the following 
generic features~\cite{Chamonetal2}:
\begin{itemize}
\item
The distribution is non-Gaussian for all $C$. 
\item
For $C\stackrel{<}{\sim} q_{ea}$ the pdf is negatively skewed and
once put into normal form it is very close to the distribution of the global 
equilibrium magnetization in the $2d$ {\sc xy} model~\cite{Holdsworth}.
It can then be approximately described by a generalized Gumbel 
form with real parameter.  
\item
In the opposite limit $C\stackrel{>}{\sim} 0$ the {\sc pdf}
is positively skewed and it does not take any 
recognizable form.
\end{itemize}
If one is interested in testing the action further one can simply use
eq.~(\ref{eq:Action-shape}) to generate, numerically, the $\psi(\vec r,t)$,
 construct the $C(\vec r,t,t_w)$ from these functions, and then compare the
{\sc pdf}s thus obtained to the ones measured, say, in a numerical
simulation of a given problem. Note that the scaling function $f_c$
also plays a role in the functional form of the {\sc pdf} of local
correlations. The same argument applies to the 
susceptibilities. 

The local coarse-grained correlations are, by construction, sums of
{\it correlated} random variables (unless $\ell \gg \xi$).  With
numerical simulations of the $3d$ {\sc ea} model~\cite{Jaubertetal}
and kinetically facilitated lattice gases~\cite{Chamonetal2} we found
that the {\sc pdf}s of correlations coarse-grained over {\it finite}
lengths $\ell$ have a functional form that resembles a generalized
Gumbel distribution characterized by a continuous parameter that
depends on $\ell/\xi$ and the value of the global correlation, $C$,
when $C\stackrel{<}{\sim} q_{ea}$ (see also~\cite{Holdsworth}).  This
fact is consistent with the discussion above and also with the
observation of Bertin and Clusel that Gumbel-like {\sc pdf}s with {\it
real} parameter characterize the statistics of sums of random
variables with {\it particular} correlations between the
elements~\cite{Bertin}. The fact that we obtain Gumbel-like
distributions then means that the correlations between the terms
in the sum are of the form needed to get this type of {\sc pdf}.

In short, the time-reparametrization scenario predicts, in its
simplest setting, that eqs.~(\ref{eq:C-shape}) and (\ref{eq:Action-shape})
fully characterise the  fluctuation of the local correlations
in the large times and coarse-graining volume  limits. 

\subsection{Two-time scaling of local functions}

As we argued in Sect.~\ref{sym-useful}, the global
time-reparametrization invariance suggests that in the ideal
asymptotic limit the slow part of the coarse-grained {\it local}
correlations and susceptibilities should scale as in
eq.~(\ref{eq:fluct-cchi}) in the ideal limit $a\ll\ell\ll\xi$. 
In practice the ideal limit is not reached and one is 
forced to work with finite correlation lengths and thus 
finite coarse-graining lengths too. The finite 
$\ell$ will then play a role and has to be taken into account. 
We now present some tests of eq.~(\ref{eq:fluct-cchi})
that are
based on the parametric representation of the 
dynamics explained in Sect.~\ref{sec:time-rep-inv}
and take into account the finite value of $\ell$.



\begin{figure}
\centerline{
\hspace{3cm}
\includegraphics[width=8cm]{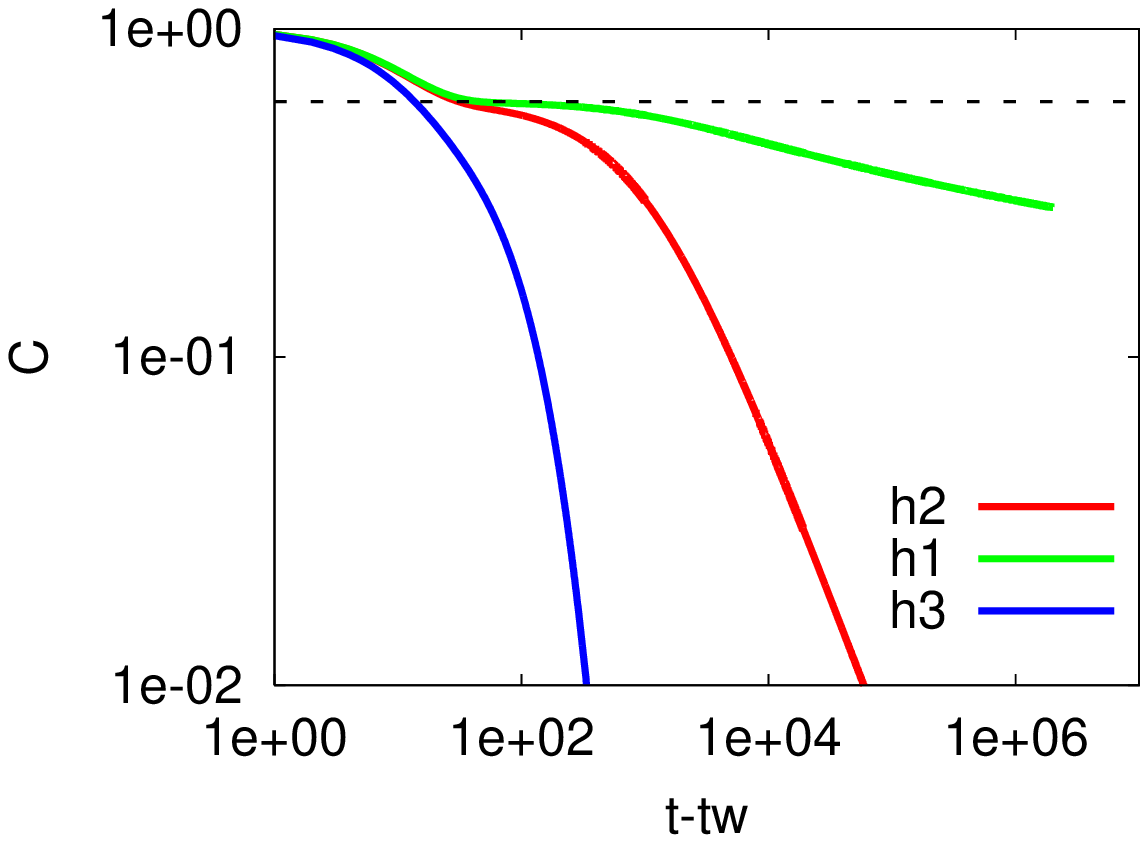}
\includegraphics[width=8cm]{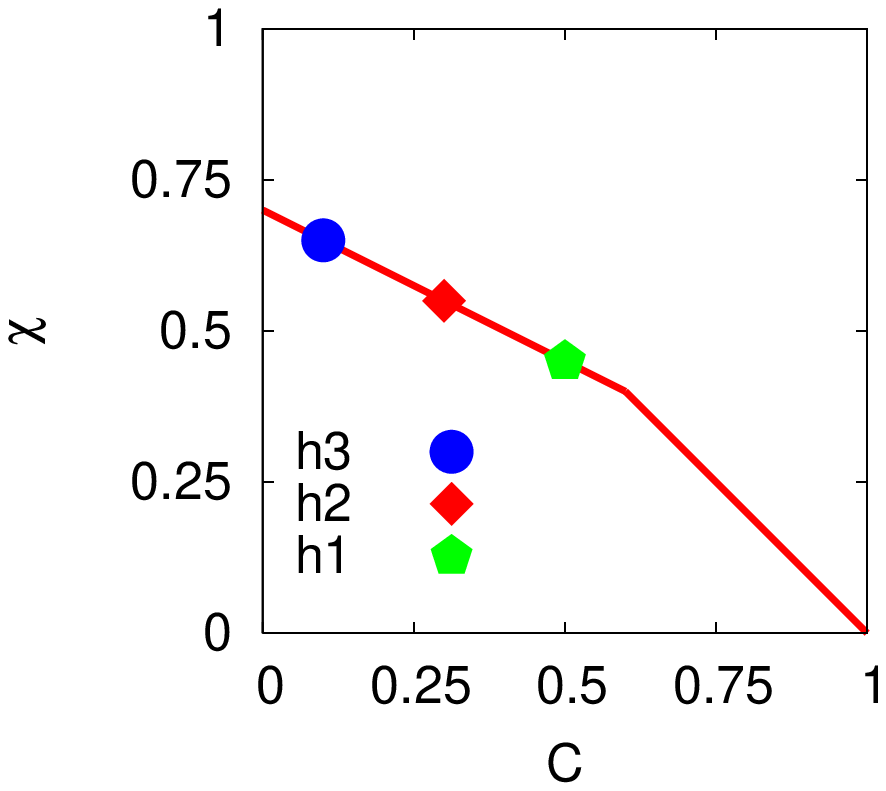}
}
\caption{Left: sketch of the decay of the correlation with the 
same stationary decay to $q_{ea}$ -- shown with a horizontal 
dashed line -- and three
choices of the scaling function $h(r_1,t)=\ln(t/t_0)$, $h(r_2,t) =t/t_0$,
 and $h(r_3,t)= e^{\ln^a(t/t_0)}$. The waiting-time is the
same in all curves. Right: the relation between the integrated
linear response against the correlation.  With a solid line, the parametric
plot for fixed and long $t_w$, using $t$ as a parameter that increases
from $t_w$ at $C=1$ to $\infty$ at $C=0$. With symbols, the three
pairs $(C_j(t,t_w), \chi_j(t,t_w))$ for the same $t_w$, a fixed
value of $t$ and $h_j(t)$ as in the left panel.}
\label{fig:sketch-hs}
\end{figure}

Let us then imagine that we compute three local coarse-grained
two-time correlations, $C_{\vec r}$, at three space points $\vec r_1$, $\vec
r_2$ and $\vec r_3$, using a given coarse-graining length, $a\ll
\ell$, and that we obtain functional forms that are characterized by
eq.~(\ref{eq:fluct-cchi}) with, say, $h(\vec r_1,t) =
\ln\left(t/t_0\right)$, $h(\vec r_2,t) = t/t_0$, and $h(\vec
r_3,t)=e^{\ln^2\left(t/t_0\right)}$, in the aging regime. In
Fig.~\ref{fig:sketch-hs} we sketch the decay of these correlations for
the same $t_w$ as a function of time-delay. The
plateau is at the same height since $q_{ea}$ as well as the the full
stationary decay are not expected to fluctuate. The external function
$f_c$ is the same in all curves. It is clear that the decay of the
three correlations  follows a different
pace, the one at $\vec r_3$ is the fastest while the relaxation at
$\vec r_2$ is the slowest.

\begin{figure}
\begin{flushright}
\includegraphics[width=8cm]{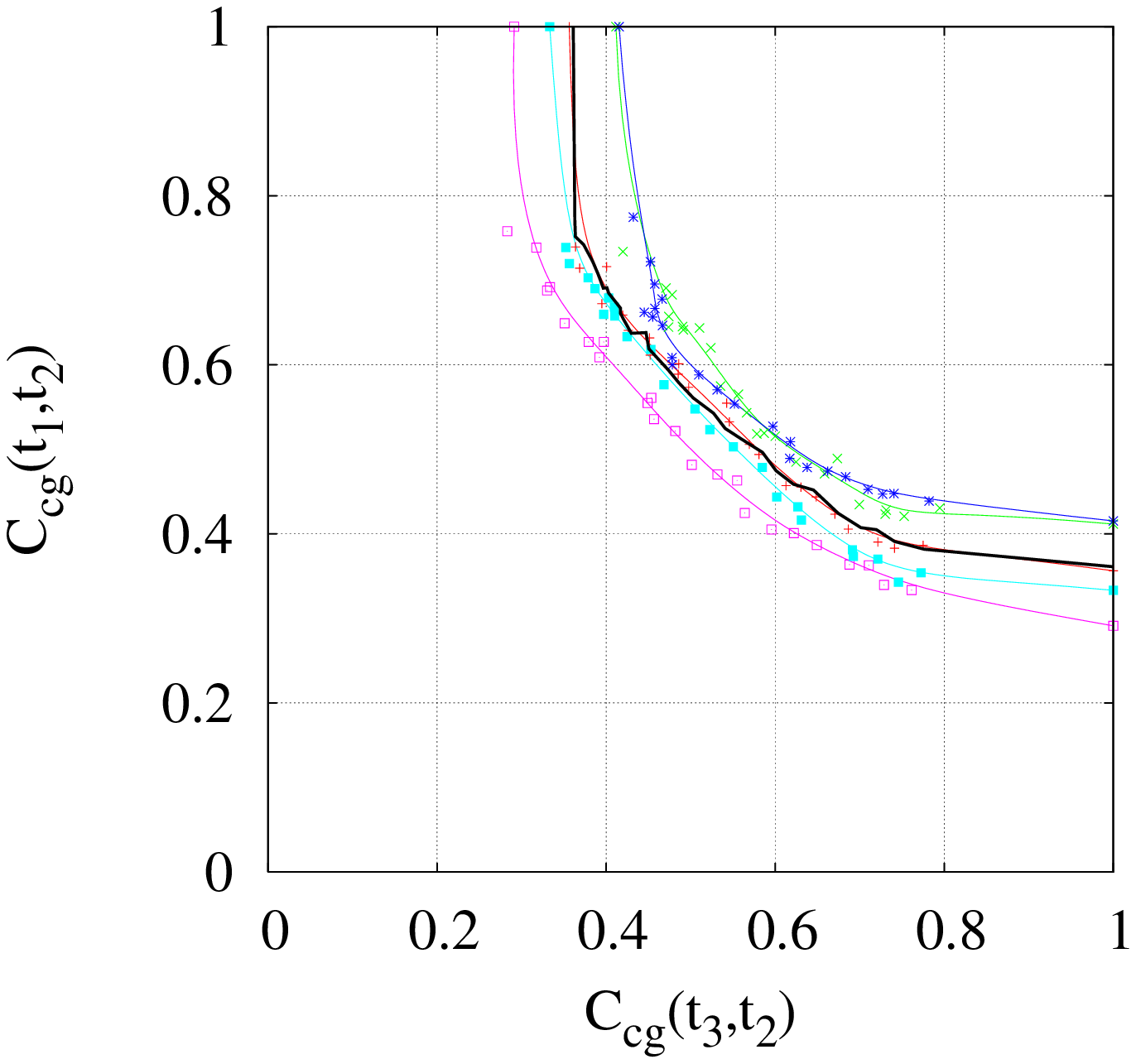}
\hspace{-1.5cm}
\includegraphics[width=8cm]{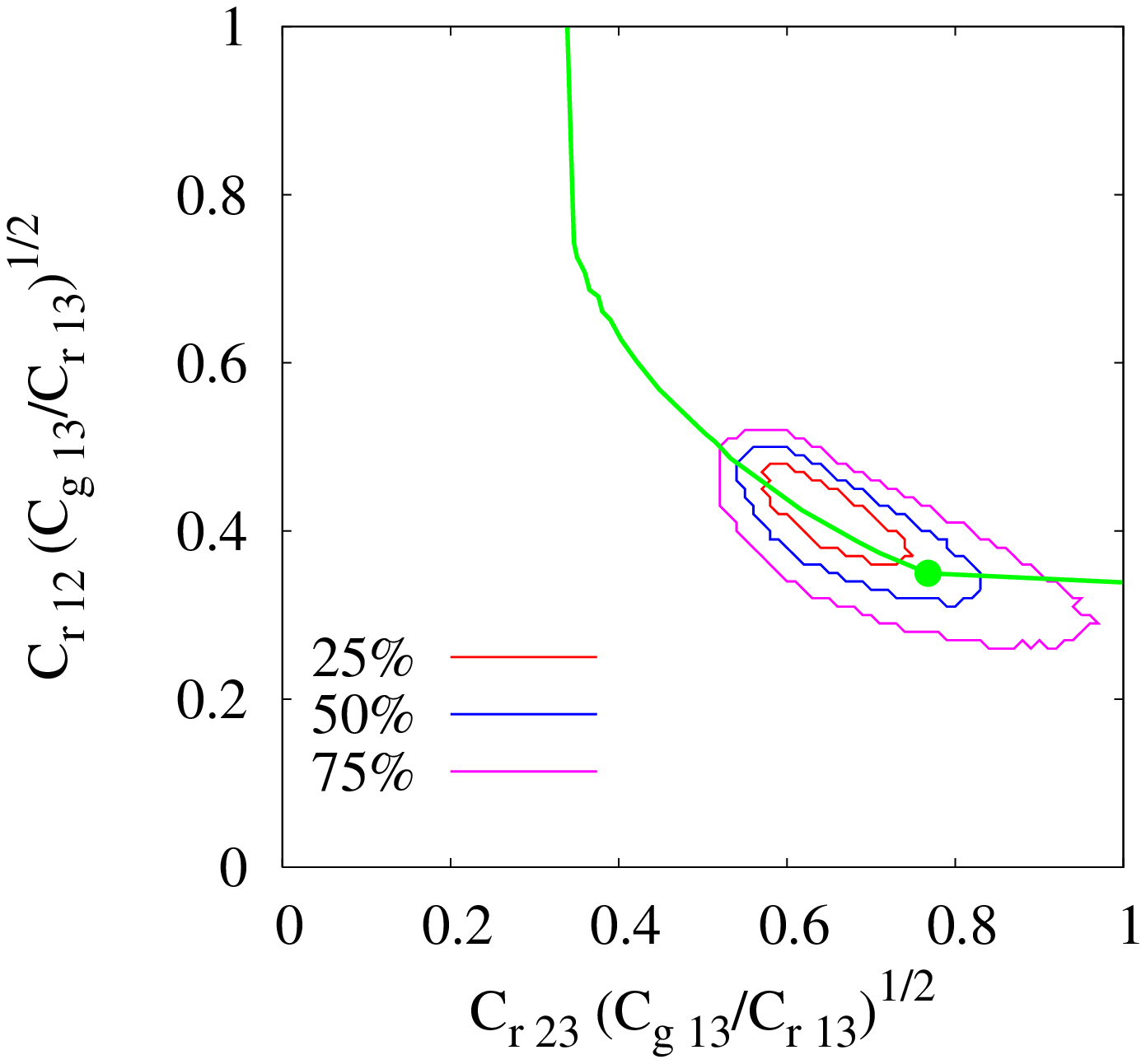}
\end{flushright}
\begin{flushright}
\includegraphics[width=8cm]{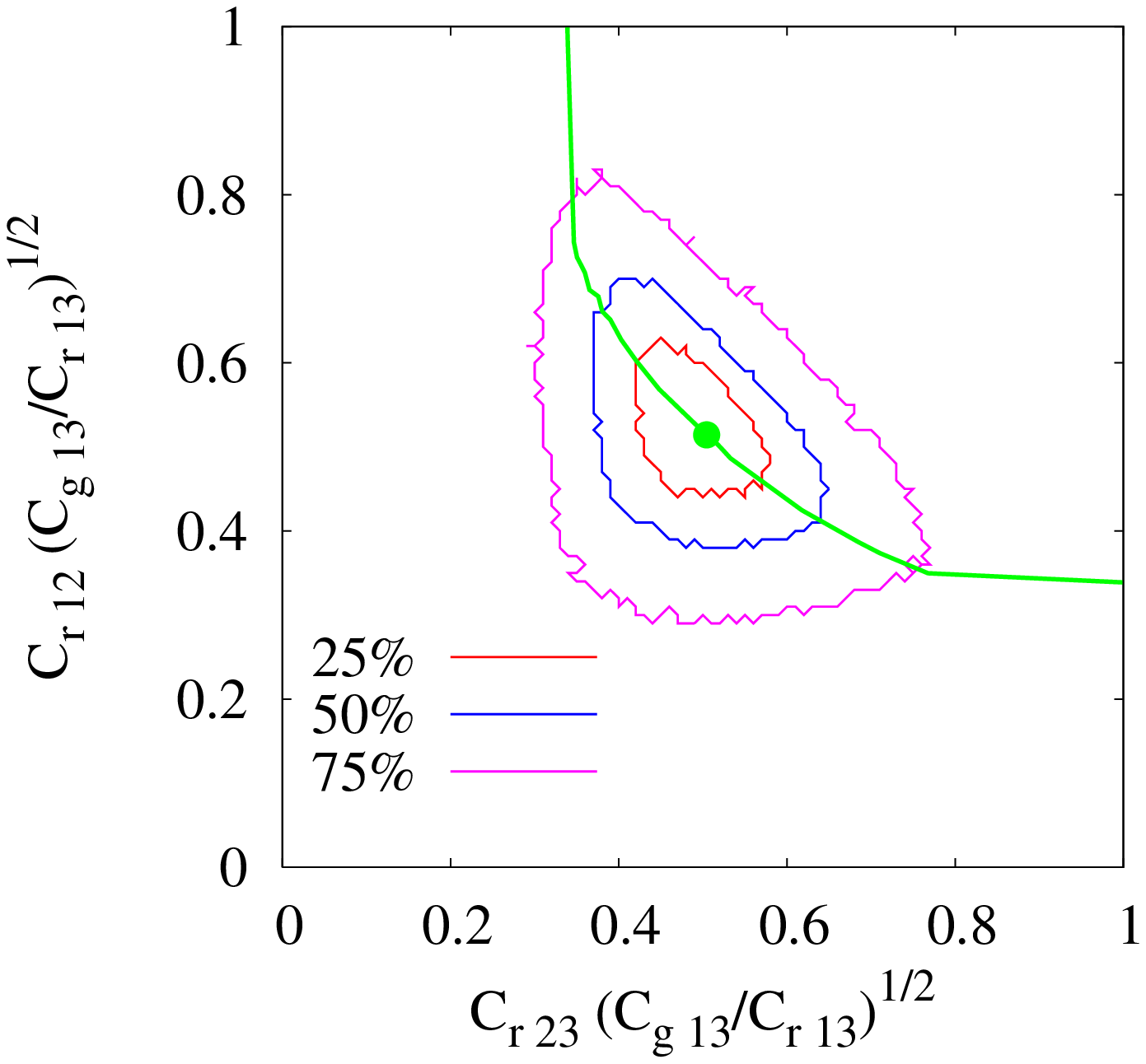}
\hspace{-1.5cm}
\includegraphics[width=8cm]{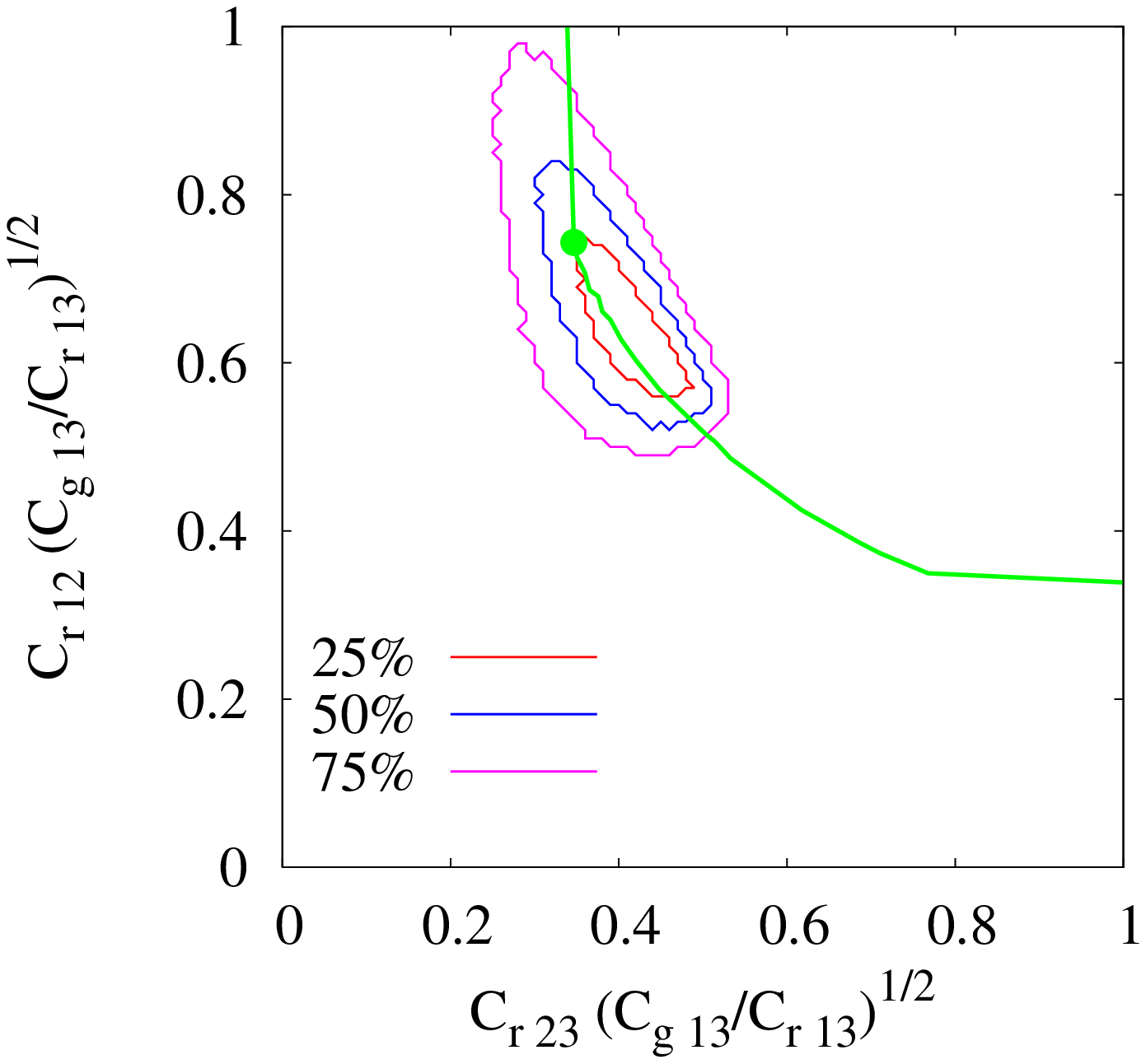}
\end{flushright}
\caption{The triangular relation in the $3d$ {\sc ea}
model. Upper left panel: the thick (black) line
represents the global $C(t_1,t_2)$ against $C(t_2,t_3)$ using $t_2$ as
a parameter varying between $t_3=5 \times 10^4$ MCs and $t_1=9 \times
10^6$ MCs, $C(t_1,t_3)\sim 0.35$ and $q_{ea}\sim 0.8$.  The curved
part is well represented by the hyperbolic form $C(t_1,t_2) \sim
q_{ea} C(t_1,t_3)/C(t_2,t_3) \sim 0.79 \times 0.35/C(t_2,t_3)$ that
corresponds to $f_c(x) \sim x^{-b}$. With different points joined with
thin lines we show the triangular relations between the local
coarse-grained correlations on five randomly chosen sites on the
lattice ($\ell=30$).  Upper right panel and lower left and right
panels: $2d$ projection of the joint probability density of $C(r;
t_1,t_2)\sqrt{C(t_1,t_3)/C(r;t_1,t_3)}$ and $C(r;
t_2,t_3)\sqrt{C(t_1,t_3)/C(r;t_1,t_3)}$ at fixed three values of the
intermediate time, $t_2=1.5\times 10^5$ MCs, $8\times 10^{5}$ MCs and
$5 \times 10^6$ MCs, respectively 
and $\ell=10$. The global
$C(t_1,t_2)$ against $C(t_2,t_3)$ using $t_2$ as a parameter is shown
with a thick green line. The green points indicate the location of
$C(t_1,t_2)$ and $C(t_2,t_3)$ for the chosen $t_2$'s. Each point in
the scatter plot corresponds to a site, $r$. The lines indicated the boundary
surrounding 25\%, 50\% and 75\% of the probability density. 
The cloud extends mostly
along the global relation as predicted by time-reparametrization
invariance. These results are taken from~\cite{Jaubertetal}.}
\label{fig:triangular-3dEA}
\end{figure}

The simplest way to put the proposal~(\ref{eq:fluct-cchi}) to the test
is to analyze the implications of eq.~(\ref{eq:fluct-cchi}) on local
triangular relations. In Sect.~\ref{sec:time-rep-inv} we showed that
two-time functions with a separation of time-scales as in
eq.~(\ref{eq:separation}) and an aging scaling as in eq.~(\ref{eq:scaling-c})
are related in a parametric way in which times disappear, see the sketch in
Fig.~\ref{fig:sketch-fig2}-left.  Equation~(\ref{eq:fluct-cchi})
implies that the local (fluctuating) two-time functions should verify
the {\it same} relation
\begin{equation}
C_{ag}(\vec r;t_1,t_3) = q_{ea} \; 
f_c\left\{ 
f^{-1}_c[C_{ag}(\vec r;t_1,t_2)/q_{ea}] \; 
f^{-1}_c[C_{ag}(\vec r;t_2,t_3)/q_{ea}] 
\right\} 
\; . 
\label{eq:triangular2} 
\end{equation} 
This is a result of the fact that time-reparametrization invariance
restricts the fluctuations to appear only in the local functions
$h(\vec r,t)$ while the function $f_c$ is locked to be the global
one
 everywhere in the sample.

A pictorial inspection of this
relation should take into account the
 fact that while the stationary
decay is not expected to fluctuate, the
 full aging relaxation and,
in particular, the minimal value of the
 local two-time functions,
$C(\vec r;t_1,t_3)$, are indeed fluctuating
 quantities. The
parametric construction on different spatial regions
 should yield
`parallel translated' curves with respect to the global
 one, as
displayed in Fig.~\ref{fig:sketch-fig2}-left.  Fluctuations in
 the
function $f_c$ would yield different functional forms in the
 curved
part of the parametric construction.  A more quantitative
 analysis
can be done by using the knowledge of $f_c$ that can be
 extracted
from the global correlation decay. Indeed, if $f_c$ is
 known, the
parametric plot
 $f_c^{-1}(C_{\vec
r12}/q_{ea})/\sqrt{f_c^{-1}(C_{\vec r13}/q_{ea})}$ against
$f_c^{-1}(C_{\vec r23}/q_{ea})/\sqrt{f_c^{-1}(C_{\vec r13}/q_{ea})}$
should
 yield a master curve identical to the global one with
different sites
 just being advanced or retarded with respect to the
global value.  
 This is another way of stating that the
 sample
ages in a heterogeneous manner, with some regions being
 younger
(other older) than the global average. (For simplicity we used a
chort-hand notation, $C_{{\vec r}\mu\nu}=C(\vec r; t_\mu,t_\nu)$
with $\mu,\nu=1,2,3$.) If
the time-reparametrization mode is indeed flat the local values
should
 lie all along this master curve in the aging regime.

The
conclusions drawn above apply in the strict $a\ll \ell \ll \xi$
limit.
 In simulations and experiments $\xi$ is finite 
 and even
rather short. Thus, $\ell$ is forced
 to also be a rather small
parameter, in which case 
 `finite size' fluctuations in $f_c$ are
also expected to 
 exist. The claim is that the latter should scale
down to zero 
 faster (in $\ell$) than the fluctuations that are
related to 
 the zero mode.

We have tested these claims in the
non-equilibrium 
 dynamics of the $3d$ {\sc ea}
spin-glass~\cite{Jaubertetal}. The 
 results are shown in
Fig.~\ref{fig:triangular-3dEA}.
 In the upper left panel we show the
global triangular relation
 (thick black line) as well as the local
one on four chosen sites. 
 The separation of time-scales is clear in
the plot. 
 The aging part is rather well described by 
 $f_c(x)
\sim x^{-b}$ and the local 
 curves are quite parallel indeed.  In
the remaining panels in Fig.~\ref{fig:triangular-3dEA} we show the
$2d$projection of the joint probability density of the site
fluctuations in the local
 coarse-grained correlations at different
chosen times $t_2$, $t_2=1.5
 \times 10^5$ MCs (upper right), $t_2=8
\times 10^5$ MCs (lower left),
 and $t_2=5 \times 10^6$ MCs (lower
right).  Taking advantage of the
 fact that $f_c(x) \sim x^{-b}$ we
use a very convenient normalization
 in which we multiply the
horizontal and vertical axes by
 $[C_{13}/C_{{\vec r}13}]^{1/2}$.
Global time-reparametrization
 invariance, expressed in
eq.~(\ref{eq:triangular2}), implies that the
 data points should
spread {\it along} the global curve indicated with
 a thick green
line in the figure. Some sites could be advanced, others
 retarded,
with respect to the global value -- shown with a point on
 the green
curve -- but all should lie mainly on the same master curve.  This
 is
indeed quite well reproduced by the simulation data in the three
cases, $C(t_1,t_2)$ close to $C(t_1,t_3)$ (upper right panel),
$C(t_1,t_2)$ close to $q_{ea}$ (lower right panel), and $C(t_1,t_2)$
far from both (lower left panel). Most of the data points tend to
follow the master curve though some fall away from it. The reason
for
 this is that eq.~(\ref{eq:triangular2}) should be strictly
satisfied
 only in the very large coarse-graining volume limit ($\ell
\gg a$)
 with $\ell/\xi \ll 1$ while we are here using $\ell=10a \sim
\xi$, see
 the discussion in Sect.~\ref{eq:two-time-length}.

The triangular relation can be used to test the fluctuations of the
local susceptibilities too. Indeed, if the separation of time scales
(\ref{eq:separation}) and the scaling (\ref{eq:scaling-c}) apply to
the global susceptibility, the local ones, after the convenient
normalization by the maximum value, should follow another master
curve, identical to the global one. Note that time-reparametrization 
invariance as we discuss it here implies that both local correlations 
and susceptibilities should be fluctuating quantities. 

Finally, notice that, in contrast to the growing correlation length
scale, there is no blatantly obvious explanation of these triangular
relations within other theoretical scenarios. These relations are
perhaps the most direct consequence of the time reparametrization
symmetry arguments, and so this prediction falls within the second
class we discussed in the introduction to this section.

\subsection{Multi-time scaling}

In general muti-time correlations are non-trivially related 
to two-time ones. One can take as an example a
generic coarse-grained connected four-time correlation. 
If this function is monotonic with respect to all times, 
and the two-point correlations scale as in 
(\ref{eq:scaling-c}) for all pairs $(t_\mu,t_\nu)$ with $\mu,\nu=1,2,3,4$, 
the four-time correlation should behave as
\begin{eqnarray}
&& C(\vec r;t_1,t_2,t_3,t_4) = g\left( \frac{h(\vec r,t_1)}{h(\vec
r,t_2)}, \frac{h(\vec r,t_2)}{h(\vec r,t_3)}, \frac{h(\vec
r,t_3)}{h(\vec r,t_4)} \right)
\label{multi-time}
\end{eqnarray}
with the same external function $g$ for all $r$, in the 
asymptotic limit in which all times are widely separated
and the corresponding two-time correlations fall below $q_{ea}$. 
Parametric constructions could be envisaged to test this 
relation. 

\subsection{Local fluctuation-dissipation relation}

The asymptotic relation between the global correlation 
and susceptibility 
\begin{equation}
\lim_{t_w\to\infty, C(t,t_w)=C} \chi(t,t_w) = \hat \chi(C)
\end{equation}
was first obtained in mean-field disordered models~\cite{Cuku1,Cuku2} and
later observed in simulations of many realistic systems
(spins and particles in interaction on finite dimensional spaces). $-(d_C\hat
\chi(C))^{-1}$ defines an effective temperature~\cite{Cukupe}.  In the
aging regime, that is to say for $C<q_{ea}$, three behaviours have
been observed in mean-field systems: in structural glass models $\hat
\chi(C)$ is linear in $C$ (solid line in the right-panel in
Fig.~\ref{fig:sketch-hs}); in spin-glass models $\hat \chi(C)$ is a
non-linear function of $C$; in coarsening systems $\hat \chi(C)$
is a constant equal to $(1-q_{ea})/T$. 

The scaling in eq.~(\ref{eq:fluct-cchi}) implies that the parametric
construction `local susceptibility against local correlation' should
fall on the master curve for the global quantities but can be advanced
or retarded with respect to the global value; again in the theoretical
limit $a\ll \ell \ll \xi$.  This behaviour is sketched in
Fig.~\ref{fig:sketch-hs}-right for the three sites whose correlations
are displayed in the left panel. The restricted relation between local
susceptibility and local correlation in eq.~(\ref{eq:fluct-cchi}) arise
from the fact that the fluctuations are due to local
reparametrizations alone and not to changes in the external functions
$f_c$ and $f_\chi$ (much as in transverse {\it vs.}  longitudinal
fluctuations in a non-linear $\sigma$-model).

 An important property of the interpretation of the fluctuation 
dissipation relation  in terms
of effective temperatures is that one expects  all observables
evolving in the same time-scale to equilibrate and hence have
the same value of the effective temperature~\cite{Cukupe} in an
asymptotic regime with slow dynamics and small entropy production.
Within the time-reparametrization invariance approach the local
effective temperature, defined from the slope of the $\hat\chi(C)$
plot, is automatically the same in the whole sample within a
correlation scale, just because the functions $f_c$ and $f_\chi$ do
not fluctuate.

\begin{figure}[h]
\centerline{
\epsfxsize=7.5cm
\epsfbox{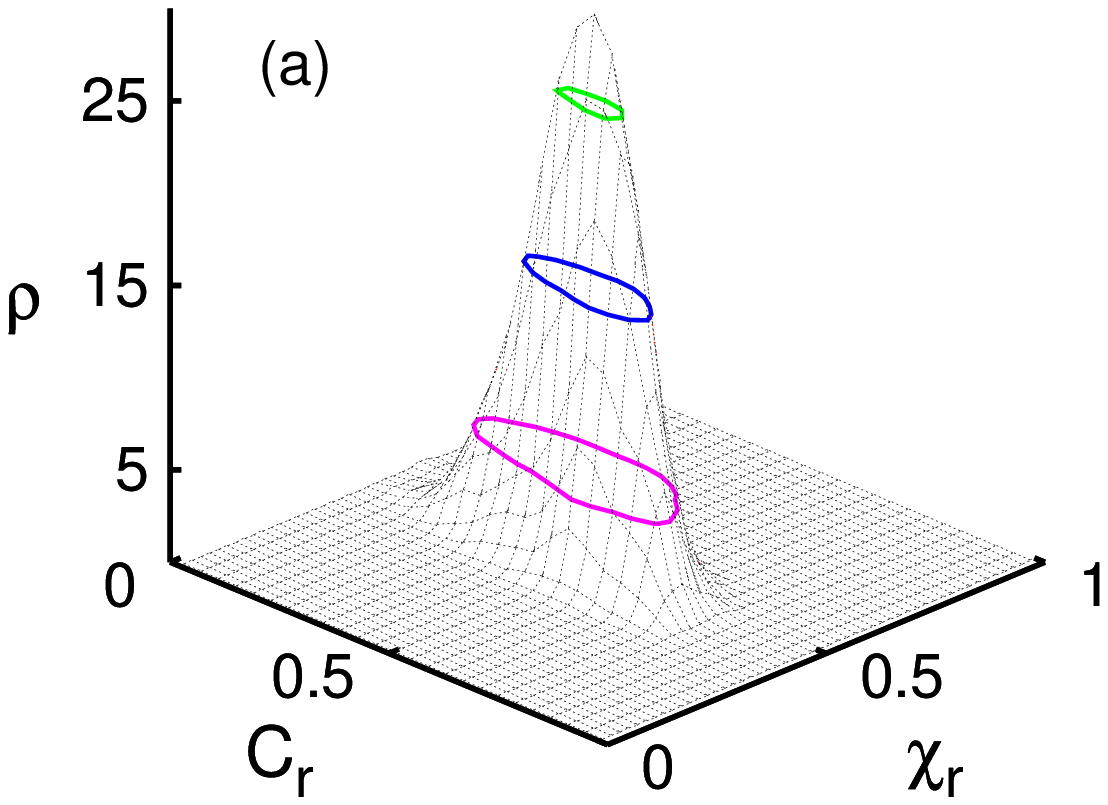}
\epsfxsize=7.5cm
\epsfbox{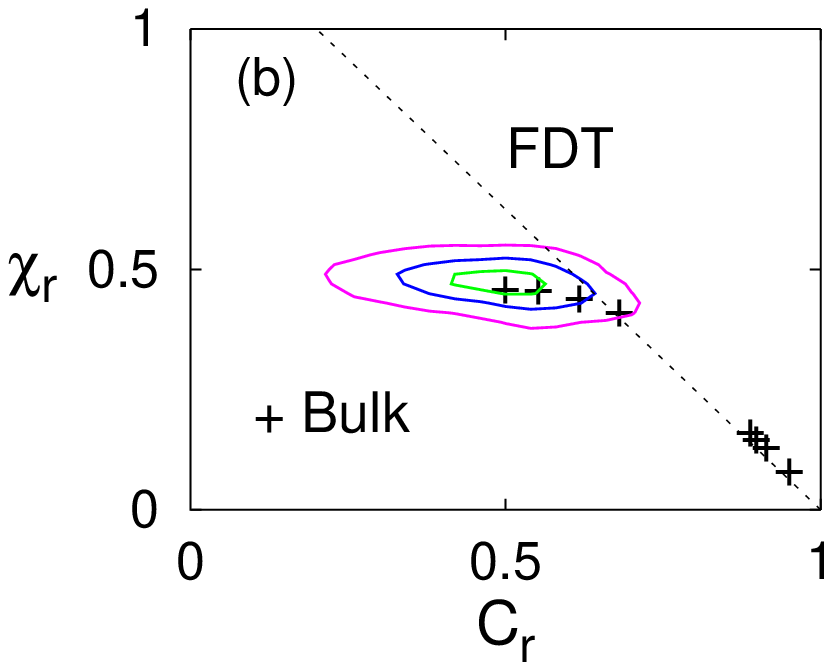}
}
\caption{(a) The joint {\sc{pdf}} $\rho(C_r,\chi_r)$ at two times
$(t_w,t)$ such that the global correlation is $C(t,t_w)=0.7<q_{\sc
ea}$ in the $3d$ {\sc ea} model. (b) Projection of three contour
levels. The crosses are the parametric construction $\hat\chi(C)$ for
several values of the total time $t$ larger than $t_w$. The dotted
straight line is {\sc fdt} at the working temperature 
$T$. These results are taken from~\cite{Castilloetal2}.}
\label{fig:dist-fdt}
\end{figure}

In Fig.~\ref{fig:dist-fdt} we show the joint 
{\sc pdf} of local correlations and susceptibilities of the
$3d$ {\sc ea} spin-glass in its glassy phase; 
the accord with the analytic prediction is
very satisfactory~\cite{Castilloetal1,Castilloetal2} with the
additional spreading away from the master curve ascribed to the fact
that $\ell$ is finite and not very different from $\xi$.

\subsection{Infinite susceptibilities}

Zero modes are intimately related to infinite 
susceptibilities. Indeed, systems with continuous symmetries are
sensitive to arbitrarily weak perturbations. In the present 
context the approximate 
global time-reparametrization invariance implies that 
one can easily change the `clock' ${\cal R}(t)$ characterizing the 
scaling of the global correlation and linear response by applying 
infinitely weak perturbations that couple to the zero mode. 
An illustration of this property is the fact that the 
aging relaxation dynamics of glassy systems is rendered stationary 
by a weak perturbing force that does not derive from a 
potential while the $\hat \chi(C)$ relation in the slow 
regime is not modified~\cite{Cukulepe}. One could envisage 
more refined tests such as applying a perturbation that imposes
different scalings on two macroscopic borders of the system and 
see how the time-reparametrization wave develops in the bulk.

\subsection{Conclusions}

In conclusion, the global time-reparametrization invariance scenario
gives a mechanism for the divergence of the correlation length $\xi$
though others have also been proposed in the literature.  There are a
number of predictions, as the parametric relations between local
coarse-grained correlations measured at different times and the local
fluctuation-dissipation relations that, to our knowledge, are not
explained by other scenarios. As regards to the easier to measure
{\sc pdf}s of local correlations the framework
is not only consistent with the scaling~(\ref{eq:pdf-scaling}) -- that
also arises from simple scaling assumptions -- but it also justifies
the non-Gaussian and Gumbel-like functional form of the {\sc pdf}s
that follows from the proposed effective action for local `ages'.
Moreover, systems with global time-reparametrization invariance 
should have as important fluctuations in the local susceptibilities
as in the local coarse-grained responses.

\section{Discussion}
\label{sec:discussion}

 We presented a summary of studies of dynamical fluctuations in
glassy
 systems that are based on the idea that, in the long time
regime, a
 global time-reparametrization invariance emerges in the
effective action describing the aging dynamics. We discussed how
this
 symmetry concretely appears in mean-field spin models, and how
it can
 be shown to emerge at the level of the action for
short-ranged spin
 glass models with quenched disorder. Two
assumptions are used to prove the global
 time-reparametrization of
the action for the short-range spin glass
 model {\it i}) causality
and unitarity, and
 {\it ii}) a separation of time scales between a
fast (or stationary)
 and a slow (or aging) relaxation, where in the
latter time translation
 invariance is broken.

That the dynamical action is symmetric under uniform, {\it i.e.},
spatially independent reparametrizations of time variables ($t\to
h(t)$) suggests that the dynamic fluctuations that cost little action
should be describable in terms of position dependent, long wavelength,
reparametrizations of the form $t\to h({\vec r},t)$. These should be the
Goldstone modes associated with breaking time-reparametrization
invariance symmetry.

 We presented predictions of this theoretical framework and
 tests
that we performed in the $3d$ Edwards-Anderson model to falsify
these predictions. Among the consequences of our theoretical
framework
 are those listed in Sect.~\ref{sec:consequences}. Some of
them find an explanation within other theories as
 well; others are
particularly related to the time reparametrization invariance
scenario. For example, a correlation length that grows in time is
associated with
 the asymptotic approach to the long-time regime in
which the symmetry
 is fully developped, and the long wavelength
modes eventually become
 massless. The existence of a growing length
is also predicted by other
 models. The functional form of the
triangular relations relating local two-time correlations between
three different times is, instead, particular to our
 framework.  

In this review we showed tests of the predictions of the global
time-reparametrization invariance scenario performed on a finite
dimensional spin-glass
model~\cite{Chamonetal1}-\cite{Jaubertetal}.  In
the past we also studied the dynamics of kinetically facilitated
models, without quenched disorder, along the same
lines~\cite{Chamonetal2}. We
 believe that, if aging dynamics is a
universal property of glassy
 systems, then these ideas should also
apply to interacting particle systems
 without quenched disorder. The
rationale is that the two assumptions
 leading to the global
time-reparametrization invariance of the
 dynamical action, namely
causality and unitarity, and a separation of
 time scales, should
also hold for other glassy systems. The former
 assumption we can
take as a fact. The second is, in a way, the
 assumption that a
glassy phase exists, even though we say nothing as
 of why it
does. As we stated in the introduction, we do not aim at the
question {\it why glasses?}, but instead we focus on the possible
universal dynamical properties once the glassy state is presented by
nature. Some of the consequences of the symmetry have already been
tested numerically in Lennard-Jones 
systems~\cite{Castillo2,Castillo1} but there is still much room 
for more detailed studies, including the analysis of local triangular 
relation between correlation and susceptibilities and tests 
of their joint behaviour.

We thus propose that the asymptotic global time-reparametrization
invariance, and the associated low action long wavelength local
reparametrizations, constitute the mechanism by which dynamic
fluctuations, that is to say heterogeneities, are generated in the
glassy state. Moreover, this mechanism may also apply, in an
approximate form, to the super-cooled liquid regime. It should just be
an approximation because in super-cooled liquids the symmetry
is not fully developed and there is then a cut-off setting the limit of the
spatial and temporal extent of the heterogeneities, in sharp constrast
to the low temperature glassy regime in which the symmetry is realized
asymptotically and fluctuations of all sizes exist. The growth and
divergence of the (two-time) dynamic correlation length defined from
the study of the space correlation of the two-time order parameter is
a manifestation of the growth and divergence of these fluctuations in
the glassy state; in contrast, such growth is interrupted in the
super-cooled liquid.

To better understand the distiction between the glassy state with
its
 asymptotic symmetry and the super-cooled state without the
fully
 developing symmetry, consider the phenomenology of dynamic
fluctuations
 as a function of temperature. Dynamic heterogeneities
in the
 super-cooled liquid phase have been identified
numerically and
experimentally~\cite{Ediger}. These are in a number of ways more
important than what is observed in a simple liquid or a solid.  In
the
 super-cooled liquid phase while the full relaxation is
stationary,
 there is still a time-scale separation with the
correlations decaying
 as a function of time-difference first rapidly
to a
 temperature-independent plateau and then slowly towards
zero. The
 latter is the structural or $\alpha$-relaxation.  The
$\alpha$
 relaxation time, $t_\alpha$, is {\it finite} in the
super-cooled
 liquid regime and it increases upon decreasing
temperature. The global parametrization chosen by the 
symmetry breaking terms in the slow regime is ${\cal R}(t) \propto
e^{-t/t_\alpha}$ and $C(t,t_w) = q_{ea} f_c(e^{-(t-t_w)/t_\alpha})$ in 
this case.

The mode-coupling approach to super-cooled liquids
is based on
 approximate dynamic equations for the relevant
correlators of
 realistic systems that are very similar to the
$p$-spin
 Schwinger-Dyson equations in the high temperature
phase~\cite{KTW,PhysicaA}.  In these equations the correlators are
already expressed as functions of the time-difference, $\tau\equiv
t-t_w$.  Close to the critical temperature the separation of
time-scales develops in the mode-coupling equations. The approximate
analysis of the $\alpha$ relaxation predicted by these equations
also
 relies on dropping the $\tau$ derivatives and approximating
the
 integrals by assuming a sharp time-scale separation. 
The remaining
asymptotic (large $\tau$) equations are invariant under
reparametrizations of $\tau$.

We then expect the local coarse-grained correlations and integrated
linear responses in the super-cooled liquid to be, to a 
first approximation, {\it stationary}
(after a sufficiently long waiting-time that goes beyond the
equilibration time) but with different finite structural relaxation
times, fluctuating about the value that characterizes the decay of the
global correlations.  This is consistent with the experimental
observation that dynamic heterogeneities in supercooled liquids seem
to have a lifetime of the order of the relaxation time. Deviations
from stationarity are not completely excluded for finite $\ell$ but they 
should be less important than in the aging low-temperature regime.

There is, however, an important difference with respect to the aging
regime, in which the equilibration time diverges and local relaxation
times or, better stated, local ages can fluctuate without limit when
$t_w\to\infty$. At high temperatures one does not expect to find
heterogeneities with arbitrary long relaxation time. Furthermore,
heterogeneities have a {\it finite} spatial extent and one can then
suppress them by using sufficiently large coarse-graining volumes. The
correlation length is stationary, $\xi(\tau)$, and it saturates in the
limit of long-time differences, $\tau\to\infty$. The saturation value,
though, increases for decreasing temperature.  From a theoretical
point of view, this picture is, in a sense, similar to the one that
describes the equilibrium paramagnetic phase in the $O(N)$ model, just
above the ordering transition temperature.

When lowering the temperature the size and life-time of the
heterogeneities increases.  A $p$-spin or
mode-coupling approach predicts that their typical size and thus
$\lim_{\tau\to\infty} \xi(\tau)$ diverge at the mode coupling
transition temperature~\cite{Silvio-length}.  In real systems the
divergence at $T_c$ is rendered smooth and $\xi$ does not strictly
diverge asymptotically.

At still lower temperatures the bulk quantities age and we expect then
to observe heterogeneous aging dynamics of the kind described in this
review, with a two-time dependent correlation length for the local
fluctuations. The heterogeneities age as well, in a `dynamic'
way. By this we mean that if a region looks older than another one
when observed on a given time-window, it can reverse its status and
look younger than the same other region when observed on a different
time-window.

The infinite susceptibility with respect to perturbations that couple
to the zero mode are illustrated by the fact that the clock of the
bulk quantities that is selected dynamically is very easy to modify
with external perturbations. A small force that does not derive from a
potential and is applied on every spin in the model renders an aging
$p$ spin model stationary~\cite{Cukulepe} while the model maintains a
separation of time-scales in which the fast scale follows the
temperature of the bath, $T$, while the slow scale is controlled by an
an effective temperature, $T_{\sc eff}>T$. In this case, the aging
system selects a time-reparametrization ${\cal R}(t)=t$ while in the
perturbed model ${\cal R}(t)=e^{-t/t_\alpha}$. Similarly, the aging of
a Lennard-Jones mixture is stopped by an homogeneous
shear~\cite{JLBarrat}. A different way to modify the
time-reparametrization that characterizes the decay of the
correlations is by using complex thermal baths~\cite{Cuku3}.

The picture that we have described applies to long times but not as
long as to enter the activated regime that we still do not know how to
characterize theoretically, not even at the bulk level. This regime
corresponds to times that scale with the system size. The success of
mean-field models, or the mode-coupling approach, in describing the
bulk dynamics of glassy systems, at least not to close to the
crossover glass temperature and at a qualitative level, allows us to
claim that these extremely long times scales are unrealistic if not 
too close to the glass transition $T_g$ even as
far as dynamic fluctuations are concerned.

The ideas discussed in this paper should not only apply to systems
that relax in a non-equilibrium manner as glasses but also to systems
with slow dynamics and a separation of time-scales that are kept out
of equilibrium with a (weak) external forcing.  Recently, there has
been much interest in the appearance of shear localization, in the
form of shear bands, in the rheology of complex fluids. Along the
lines here described it would be very interesting to analyze the
fluctuations in the local reparametrizations in the fluidized shear
band and the `jammed' glassy band.

The analytic treatment of mean-field {\it quantum} glassy systems follows 
similar steps to the ones presented here. The Schwinger-Keldysh approach 
replaces the Martin-Siggia-Rose one but these formalisms are very similar 
indeed. The analytic solution to the dynamic equations in the limit in 
which the coupling to the environment is weak also uses the fact that 
the dynamics in the aging regime is very slow. The approximate
equations then become time-reparametrization invariant. One can 
then expect that similar dynamic fluctuations arise in glassy problems
in which quantum fluctuations are important. Moreover the proof of 
global time-reparametrization invariance for spin-glasses has been presented
directly in the quantum formalism. Novel experimental techniques may be apt 
to study dynamic heterogeneities in glasses when quantum fluctuations 
are important.

We expect to find similar fluctuations using
finite size systems and examining the behaviour of the mesoscopic
run-to-run fluctuations of the global correlations. These may be easier
to measure experimentally using mesoscopic systems.

Importantly enough, global time reparametrization invariance does not 
develop in all models with slow and aging correlation functions. 
The O(N) ferromagnetic model in the large $N$ limit is a case in 
which global time-reparametrization invariance is reduced to just 
scale invariance~\cite{Chamonetal3}.

Last, but not least, the approach based on reparametrization
invariance suggests that it may be possible to search for universality
in glassiness. A Ginzburg-Landau theory for phase
transitions captures universal properties that are independent of the
details of the material. It is symmetry that defines the universality
classes. For example, one requires rotational invariance of the
Ginzburg-Landau action when describing ferromagnets. Time reparametrization
invariance may be the underlying symmetry that must be satisfied by
the Ginzburg-Landau action of all glasses.  What would determine if a
system is glassy or not? We are tempted to say the answer is if the
symmetry is generated or not at long times. Knowing how to describe
the universal behavior may tell us all the common properties of 
glasses, but surely it will not allow us to make non-universal
predictions, such as what is the glass transition temperature for a
certain material, or whether the material displays glassy behavior at
all. This quest for universality is a very interesting theoretical
scenario that needs to be confronted.

\vspace{0.5cm}

We have tried to state as clearly as possible the implications 
of our proposal. The full project is not yet complete since several 
questions about its limitations remain open. 
A number of issues should be addressed are:

(i) From a phenomenological point of view, to perform strong tests of
the global time-reparametrization invariance scenario in molecular
dynamic simulations~\cite{Parisi,Valluzzietal,Vollmayr} of 
realistic glassy systems and experiments~\cite{Sergio2}-\cite{Makse}. More
precisely, the triangular relations between local coarse-grained
correlations and the local fluctuation dissipation relations should
be analysed to give support or else falsify this conjecture.

(ii) From an analytic point of view, to derive the effective action 
for local reparametrizations for glassy models with and without 
quenched disorder. We are currently 
working on this project in collaboration with S. Franz. One idea
is to study the $p$ spin disordered model with Kac long-range interaction.
ANothe one is to study the symmetries of the dynamic action associated 
with the Dean-Kawasaki equation for the evolution of the density of 
a system of particles in interaction.

(iv) 
 From a mixed analytic and numerical point of view
to  analyse fluctuations in models with global 
aging dynamics of different type. To this end, one can 
study dynamic fluctuations in simple coarsening systems in 
finite dimensions. In these cases the morphology of domains 
can be characterized in great detail~\cite{Arenzonetal}. We 
could, in principle, understand the fluctuations in the local 
correlations and linear responses from a microscopic point of 
view. Whether these are similar or different to the ones 
in other glassy problems is still to be established and the outcome 
of this study could clarify the relevance of the value of the 
effective temperature in determining the characteristics of 
the dynamic fluctuations. 

(v) In the same line as the above, the analysis of fluctuations of
elastic lines in the presence of impurities could help us understanding
the coarsening phenomenon but also the role of diffusion
that superimposes in these cases to the aging dynamics~\cite{Bustingorryetal1}. 

These are just a few questions posed by this proposal that are still
not answered.  

\vspace{2cm}
\noindent 
\underline{Acknowledgments}
\vspace{0.5cm}
 
 We thank our collaborators 
 J. J. Arenzon,
C. Aron, 
 A. J. Bray, S. Bustingorry, H. E. Castillo,
P. Charbonneau, D. Dom\'{\i}nguez, 
 J. L. Iguain, L. D. C. Jaubert,
M. P. Kennett, M. Picco, D. R. Reichman, 
 M. Sellitto, A. Sicilia
and H. Yoshino. 
 
 We also wish to especially thank 
 L. Berthier,
G. Biroli, J-P Bouchaud, D. S. Dean, 
 G. Fabricius, T. Grigera,
E. Fradkin, S. Franz, J. Kurchan, H. Makse, 
 D. Stariolo and
L. Valluzzi for very helpful discussions. 
 We acknowledge financial
support from NSF-CNRS INT-0128922, NSF DMR-0305482, DMR 0403997 and
PICS 3172.  LFC is a member of Institut Universitaire de France. LFC
thanks the Newton Institute at the University of Cambridge, ICTP
at Trieste, and Universidad Nacional de Mar del Plata, Argentina,
CC the LPTHE at Jussieu, Paris, France, 
and LFC and CC the Aspen Center for Physics for hospitality
where part of this work has been carried out.

\end{document}